\documentclass{paper}

\usepackage{graphicx}
\usepackage{amsmath,epsfig,amssymb,mathrsfs}
\usepackage{algorithmic}
\usepackage{cite}
\usepackage{amsthm}
\usepackage{arydshln}
\usepackage{amsopn}
\usepackage{color}
\usepackage{verbatim}
\usepackage{amsopn}
\usepackage{amssymb}
\usepackage{algorithm}
\usepackage{enumerate}

\newtheorem{theorem}{Theorem}[section]
\newtheorem{lemma}[theorem]{Lemma}
\newtheorem{definition}[theorem]{Definition}

\newtheorem{corollary}[theorem]{Corollary}

\newcommand{\Ttau}{\tilde{\tau}}
\newcommand{\cO}{\mathcal{O}}

\newcommand{\bE}{\mathbb{E}}

\newcommand{\bP}{\mathbb{P}}

\begin{document}
\title{Parameterless Optimal Approximate Message Passing}
\author{Ali Mousavi, Arian Maleki, Richard G. Baraniuk}
\maketitle
\begin{abstract} 
Iterative thresholding algorithms are well-suited for high-dimensional problems in sparse recovery and compressive sensing. The performance of this class of algorithms depends heavily on the tuning of certain threshold parameters. In particular, both the final reconstruction error and the convergence rate of the algorithm crucially rely on how the threshold parameter is set at each step of the algorithm. In this paper, we propose a parameter-free approximate message passing (AMP) algorithm that sets the threshold parameter at each iteration in a fully automatic way without either having an information about the signal to be reconstructed or needing any tuning from the user. We show that the proposed method attains both the minimum reconstruction error and the highest convergence rate. Our method is based on applying the Stein unbiased risk estimate (SURE) along with a modified gradient descent to find the optimal threshold in each iteration. Motivated by the connections between AMP and LASSO, it could be employed to find the solution of the LASSO for the optimal regularization parameter. To the best of our knowledge, this is the first work concerning parameter tuning that obtains the fastest convergence rate with theoretical guarantees. 
\end{abstract}

\section{Introduction}
\subsection{Motivation}
Compressed sensing (CS) is concerned with the problem of recovering a sparse vector $x_o \in \mathbb{R}^N$ from a noisy undersampled set of linear observations acquired via $y= Ax_o + w$, where $w \in \mathbb{R}^n$ and $A \in \mathbb{R}^{n \times N}$ denote the noise and measurement matrix, respectively. The success of CS in many applications has encouraged researchers to apply it to ambitious high-dimensional problems such as seismic signal acquisition and MRI. In such applications, the acquisition step requires simple modifications in the current technology. However, the recovery phase is challenging as the recovery algorithms are usually computationally demanding. Iterative thresholding algorithms have been proposed as a simple remedy for this problem. Among these iterative thresholding algorithms, approximate message passing (AMP) has recently attracted attention for both its simplicity and its appealing asymptotic properties. Starting with the initial estimate $x^0= 0$, AMP employs the following iteration:
\begin{align} \label{equ:AMP}
x^{t+1} &=\eta(x^t+A^*z^t;\tau^{t-1}),  \nonumber \\
z^t &=y-Ax^t+ \langle \eta'(x^{t-1}+A^*z^{t-1};\tau^{t-1}) \rangle.
\end{align}
Here $\eta$ is the soft-thresholding function that is applied component-wise to the elements of a vector, and $\tau^t$ is called the threshold parameter at iteration $t$.  $x^t \in \mathbb{R}^N$ and $z^t \in \mathbb{R}^n$ are the estimates of signal $x_o$ and the residual $y-Ax_o$ at iteration $t$, respectively. Finally, $A^*$ is the transpose of the matrix $A$, and $\eta'$ is the derivative of the soft thresholding function. We will describe the main properties of AMP in more detail in Section \ref{sec:amp1}.

One of the main issues in using iterative thresholding algorithms in practice is the tuning of their free parameters. For instance, in AMP one should tune $\tau^1, \tau^2, \ldots$ properly to obtain the best performance. The $\tau^t$ have a major impact on the following aspects of the algorithm:
\begin{itemize}
\item [] (i) The final reconstruction error, $\lim_{t \rightarrow \infty} \|{x}^t- x_o\|_2^2/N$. Improper choice of $\tau^t$ could lead the algorithm not to converge to the smallest final reconstruction error.
\item[] (ii) The convergence rate of the algorithm to its final solution. A bad choice of $\tau^t$ leads to extremely slow convergence of the algorithm. 

\end{itemize}
Ideally speaking, one would like to select the parameters in a way that the final reconstruction error is the smallest while simultaneously the algorithm converges to this solution at the fastest achievable rate. Addressing these challenges seem to require certain knowledge about $x_o$. In particular, it seems that for a fixed value of $\tau$, $\lim_{t \rightarrow \infty} \|x^t- x_o\|_2^2$ depends on $x_o$. Therefore, the optimal value of $\tau$ depends on $x_o$ as well. This issue has motivated researchers to consider the least favorable signals that achieve the maximum value of the mean square error (MSE) for a given $\tau$ and then tune $\tau^t$ to obtain the minimum MSE for the least favorable signal \cite{DoMaMoNSPT, MaDo09sp, MaAnZaBa13}. These schemes are usually too pessimistic for practical purposes.   \\

The main objective of this paper is to show that the properties of the AMP algorithm plus the high dimensionality of the problem enable us to set the threshold parameters $\tau^t$ such that 
(i) the algorithm converges to its final solution at the highest achievable rate, and (ii) the final solution of the algorithm has the minimum MSE that is achievable for AMP with the optimal set of parameters. 

 The result is a parameter-free AMP algorithm that requires no tuning by the user and at the same time achieves the minimum reconstruction error and highest convergence rate. The statements claimed above are true asymptotically as $N \rightarrow \infty$. However, our simulation results show that the algorithm is successful even for medium problem sizes such as $N = 1000$.  We will formalize these statements in Sections \ref{subsec:tune} and \ref{sec:optAMP}. 
 
%
%

\subsection{Implications for LASSO}
One of the most popular sparse recovery algorithms is the LASSO, which minimizes the following cost function:
\[
\hat{x}^{\lambda} =\arg \min_x \frac{1}{2} \|y-Ax\|_2^2 + \lambda \|x\|_1. 
\]
$\lambda \in (0, \infty)$ is called the regularization parameter. The optimal choice of this parameter has a major impact on the performance of LASSO. It has been shown that the final solutions of AMP with different threshold parameters corresponds to the solutions of the LASSO for different values of $\lambda$ \cite{mousavi2013asymptotic,DoMaMo09, DoMaMoNSPT, BaMo10, BaMo11}

 This equivalence implies that if the parameters of the AMP algorithm are tuned ``optimally'', then the final solution of AMP corresponds to the solution of LASSO for the optimal value of $\lambda$, i.e., the value of $\lambda$ that minimizes the MSE, $\|\hat{x}^{\lambda} - x_o \|_2^2/N$. Therefore, finding the optimal parameters for AMP automatically provides the optimal parameters for LASSO as well. 

\subsection{Related Work}\label{sec:relwork}

We believe that this is the first paper to consider the problem of setting threshold parameters to obtain the fastest convergence rate of an iterative thresholding algorithm. Several other papers that consider various threshold-setting strategies to improve the convergence rate \cite{HaYiZh, becker2011templates}. However, these schemes are based on heuristic arguments and lack theoretical justification.

Optimal tuning of parameters to obtain the smallest final reconstruction error has been the focus of major research in CS, machine learning, and statistics. The methods considered in the literature fall into the following three categories: 
\begin{itemize}
\item[(i)] The first approach is based on obtaining an upper bound for the reconstruction error and setting the parameters to obtain the smallest upper bound. For many of the algorithms proposed in the literature, there exists a theoretical analysis based on certain properties of the matrix, such as RIP \cite{candes2006stable,candes2008restricted}, Coherence \cite{donoho2003optimally}, and RSC \cite{bickel2009simultaneous}. These analyses can potentially provide a simple approach for tuning parameters. However, they suffer from two issues: (i) Inaccuracy of the upper bounds derived for the risk of the final estimates usually lead to pessimistic parameter choices that are not useful for practical purposes. (ii) The requirement of an upper bound for the sparsity level \cite{BlDa08, Maleki09}, which is often not available in practice. 

\item[(ii)] The second approach is based on the asymptotic analysis of recovery algorithms. The first step in this approach is to employ asymptotic settings to obtain an accurate estimate of the reconstruction error of the recovery algorithms. This is done through either pencil-and-paper analysis or computer simulation. The next step is to employ this asymptotic analysis to obtain the optimal value of the parameters. This approach is employed in \cite{DoMaMoNSPT}. The main drawback of this approach is that the user must know the signal model (or at least an upper bound on the sparsity level of the signal) to obtain the optimal value of the parameters. Usually, an accurate signal model is not available in practice, and hence the tuning should consider the least favorable signal that leads to pessimistic tuning of the parameters.  

\item[(iii)] The third approach involves model selection ideas that are popular in statistics. For a review of these schemes refer to Chapter 7 of \cite{trevor2001elements}. Since the number of parameters that must be tuned in AMP is too large (one parameter per iteration), such schemes are of limited applicability. However, as described in Section \ref{ssec:ampintui}, the features of AMP enable us to employ these techniques in certain optimization algorithms and tune the parameters efficiently. 
\end{itemize}
 
Rather than these general methods, other approaches to skip the parameter tuning of AMP is proposed in \cite{SchSel11, Schniter10, ViSc12, KamRang12}. These approaches are inspired by the Bayesian framework; a Gaussian mixture model is considered for $x_o$, and then the parameters of that mixture are estimated at every iteration of AMP by using an expectation-minimization technique \cite{ViSc12}. While these schemes perform well in practice, there is no theoretical result to confirm these observations. A first step toward a mathematical understanding of these methods is taken in \cite{KamRang12}.

\subsection{Notation}
We use calligraphic letters like $\mathcal{A}$ to denote the sets and capital letters are used for both the matrices and random variables. $\bE$, $\bP$, and $\bE_X$ are symbols used for expected value, probability measure, and expected value with respect to random variable $X$, respectively. For a vector $x\in \mathbb{R}^n$ we denote by $\|x\|_0 = |\{i \ : \  |x_i| \neq 0 \} |$ and $\| x\|_p \triangleq (\sum |x_i|^p)^{1/p}$ the $\ell_0$ and $\ell_p$ norms, respectively. Either for a variable or a matrix we may use notion like $x_o(N)$ and $A(N)$ in order to show the dependency on the ambient dimension $N$. $\mathbb{I}(\cdot)$ denotes the indicator function and finally, $\mathcal{O}(\cdot)$ and $o(\cdot)$ are denoting ``big O" and ``small O" notations, respectively.
%

\section{Our approach for tuning AMP} \label{sec:amp1}

\subsection{Intuitive explanation of the AMP features}\label{ssec:ampintui}
\begin{figure}[t!]
\centering
\includegraphics[width= 9cm]{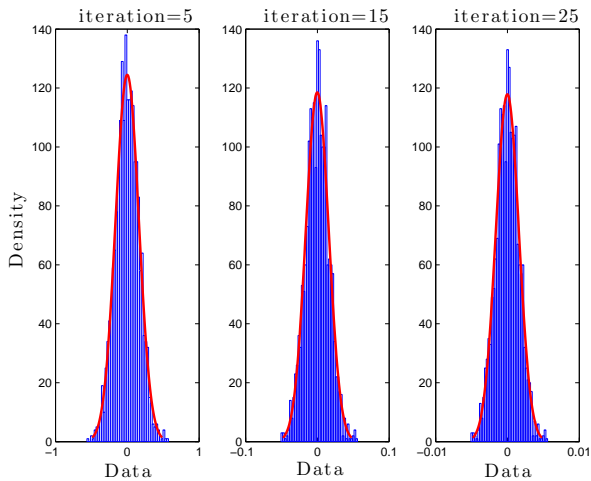}
\caption{Blue bars show the histogram of $v^t$ which are approximately Gaussian. The red curves displays the best Gaussian fit. In this experiment $N=2000$, $\delta=0.85$, $\rho=0.2$ and the measurement matrix is a random Gaussian noise.}
\label{fig:gaussiancheck}
\end{figure}

In this section, we summarize some of the main features of AMP intuitively. The formal exposition of these statements will be presented in Section \ref{sec:formalstateAMP}. Consider the iterations of AMP defined in \eqref{equ:AMP}. Define $\tilde{x}^t \triangleq x^t+ A^*z^t$ and $v^t \triangleq \tilde{x}^t-x_o$. We call $v^t$ the noise term at the $t^{\rm th}$ iteration. Clearly, at every iteration AMP calculates $\tilde{x}^t$. In our new notation this can be written as $x_o + v^t$.  If the noise term $v^t$ has iid zero-mean Gaussian distribution and is independent of $x_o$, then we can conclude that at every iteration of AMP the soft thresholding is playing the role of a denoiser. The Gaussianity of $v^t$, if holds, will lead to deeper implications that will be discussed as we proceed. To test the validity of  this noise model we have presented a simulation result in Figure \ref{fig:gaussiancheck}. This figure exhibits the histogram of $v^t$ overlaid with its Gaussian fit for a CS problem. It has been proved that the Gaussian behavior we observe for the noise term is accurate in the asymptotic settings \cite{DoMaMo09, DoMaMoNSPT, BaMo10}. We will formally state this result in Section \ref{sec:formalstateAMP}.

In most calculations, if $N$ is large enough that we can assume that $v^t$ is iid Gaussian noise. This astonishing feature of AMP leads to the following theoretically and practically important implications:

\begin{itemize}
\item[(i)] The MSE of AMP, i.e., $\frac{\|x^t-x_o\|_2^2}{N}$ can be theoretically predicted (with certain knowledge of $x_o$) through what is known as state evolution (SE). This will be described in Section \ref{sec:formalstateAMP}.
\item[(ii)]  The MSE of AMP can be estimated through  the Stein unbiased risk estimate (SURE). This will enable us to optimize the threshold parameters.  This scheme  will be described in the next section. 
\end{itemize}

\subsection{Tuning scheme}\label{ssec:tuneint}
\begin{figure}[t!]
\centering
\includegraphics[width= 11cm]{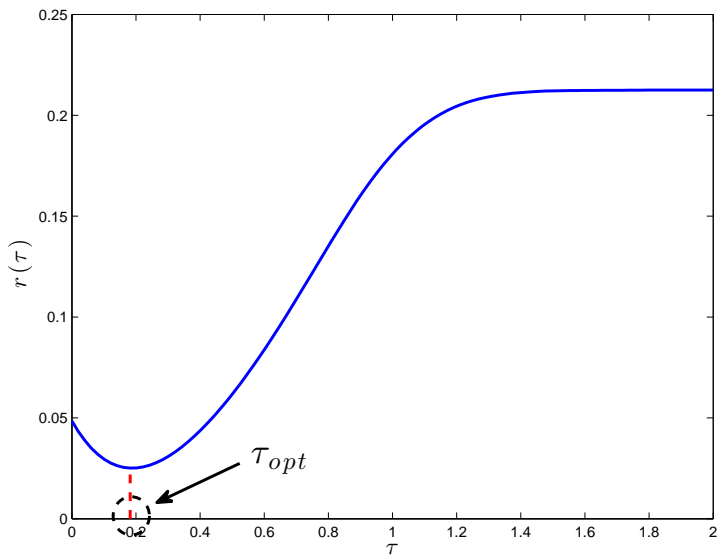}
\caption{Risk function $r(\tau, \sigma)$ as a function of the threshold parameter $\tau$. $x_o \in \mathbb{R}^N$ is a $k$-sparse vector where $N=2000$ and $k=425$. In addition, $\sigma = 0.2254$ where $\sigma$ is the standard deviation of the noise in the model $\tilde{x}^t =x_o^t+v^t$.}
\label{fig:riskandtau}
\end{figure}

In this section we assume that each noisy estimate of AMP, $\tilde{x}^t$ , can be modeled as $\tilde{x}^t = x_o + v^t$, where $v^t$ is an iid Gaussian noise as claimed in the last section, i.e., $v^t \sim N(0, \sigma_t^2 I)$, where $\sigma_t$ denotes the standard deviation of the noise. The goal is to obtain a better estimate of $x_o$. Since $x_o$ is sparse, AMP  applies the soft thresholding to obtain a sparse estimate ${x}^t = \eta(\tilde{x}^t ; \tau^t)$. The main question is how shall we set the threshold parameter $\tau^t$? To address this question first define the risk (MSE) of the soft thresholding estimator as
\[
r(\tau ; \sigma) = \frac{1}{N} \mathbb{E} \|\eta(x_o + \sigma u ; \tau) - x_o  \|_2^2,
\]
where $u \sim N(0, I)$. Figure \ref{fig:riskandtau} depicts $r(\tau, \sigma)$ as a function of $\tau$ for a given signal $x_o$ and given noise level $\sigma$. In order to maximally reduce the MSE we have to set $\tau$ to $\tau_{opt}$ defined as 
\[
\tau_{opt} = \arg \min_{\tau} r(\tau).
\]

There are two major issues in finding the optimizing parameter $\tau_{opt}$: (i) $r(\tau, \sigma)$ is a function of $x_o$ and hence is not known. (ii) Even if the risk is known then it seems that we still require an exhaustive search over all the values of $\tau$ (at a certain resolution) to obtain $\tau_{opt}$. This is due to the fact that $r(\tau, \sigma)$ is not necessarily a well-behaved function, and hence more efficient algorithms such as gradient descent or Newton method do not necessarily converge to $\tau_{opt}$.  

Let us first discuss the problem of finding $\tau_{opt}$ when the risk function $r(\tau, \sigma)$ and the noise standard deviation $\sigma$ are given. In a recent paper we have proved that $r(\tau, \sigma)$ is a quasi-convex function of $\tau$ \cite{mousavi2013asymptotic}. Furthermore, the derivative of $r(\tau, \sigma)$ with respect to $\tau$ is {\em only} zero at $\tau_{opt}$. In other words, the MSE does not have any local minima except for the global minima. Combining these two facts we will prove in Section \ref{sec:gdidealrisk} that if the gradient descent algorithm is applied to  $r(\tau, \sigma)$, then it will converge to $\tau_{opt}$. The ideal gradient descent is presented in Algorithm \ref{alg:idealgrad}. We call this algorithm the ideal gradient descent since it employs $r(\tau, \sigma)$ that is not available in practice. 

The other issue we raised above is that in practice the risk (MSE) $r(\tau, \sigma)$ is not given. To address this issue we employ an estimate of $r(\tau,\sigma)$ in the gradient descent algorithm. The following lemma known as Stein's unbiased risk estimate (SURE) \cite{donoho1995adapting} provides an unbiased estimate of the risk function:
\begin{lemma} {\rm \cite{St81Es}} \label{lem:stein}
Let $g(\tilde{x})$ denote the denoiser. If $g$ is weakly differentiable, then 
\begin{equation}\label{eq:stein}
\mathbb{E} \| g(\tilde{x}) - x_o\|^2/N = \mathbb{E} \|g(\tilde{x}) - \tilde{x}\|_2^2/N-  \sigma^2 + 2 \sigma^2  \mathbb{E} (\mathbf{1}^T (\nabla g(\tilde{x})-\mathbf{1}))/N,
\end{equation}
where $\nabla g(\tilde{x})$ denotes the the gradient of $g$ and $\mathbf{1}$ is an all one vector. 
\end{lemma} 

This lemma provides a simple unbiased estimate of the risk (MSE) of the soft thresholding denoiser:
\[
\hat{r}(\tau, \sigma) = \|\eta(\tilde{x}; \tau) - \tilde{x}\|^2/N-  \sigma^2 + 2 \sigma^2  (\mathbf{1}^T ( \eta'(\tilde{x};\tau)-\mathbf{1}))/N,
\]
We will study the properties of $\hat{r}(\tau, \sigma)$ in Section \ref{sec:empriskth} and we will show that this estimate is very accurate for high dimensional problems. Furthermore, we will show how this estimate can be employed to provide an estimate of the derivative of $r (\tau, \sigma)$ with respect to $\tau$. Once these two estimates are calculated, we can run the gradient descent algorithm for finding $\tau_{opt}$. We will show that the gradient descent algorithm that is based on empirical estimates converges to $\hat{\tau}_{opt}$, which is ``close'' to $\tau_{opt}$ and converges to $\tau_{opt}$ in probability as $N \rightarrow \infty$. We formalize these statements in Section \ref{sec:graddesth}. 

\begin{algorithm}[!t]                         
\label{alg:GDRisk}                     
\begin{algorithmic}                    
\REQUIRE $r(\tau),\epsilon,\alpha$
\ENSURE $\arg \min_{\tau} r(\tau)$ 
\WHILE {$r'(\tau)>\epsilon$}
	\STATE $\tau=\tau - \alpha \Delta_{\tau}$
\ENDWHILE
\end{algorithmic}
\caption{Gradient descent algorithm when the risk function is exactly known. The goal of this paper is to approximate the iterations of this algorithm. }
\label{alg:idealgrad}
\end{algorithm}

\subsection{Roadmap}

Here is the organization of the rest of the paper. Section \ref{subsec:tune} considers the tuning of the threshold parameter for the problem of denoising by soft thresholding. Section \ref{sec:optAMP} connects the results of optimal denoising discussed in Section \ref{subsec:tune} with the problem of optimal tuning of the parameters of AMP. Section \ref{sec:proofs} includes the proofs of our main results. Section \ref{sec:simul} presents our simulation results. Finally, Section \ref{sec:conclusion} summarizes the contributions of the paper and outlines several open directions for the future research.

\section{Optimal parameter tuning for denoising problems}\label{subsec:tune}
This section considers the problem tuning the threshold parameter in the soft-thresholding denoising scheme. Section \ref{sec:optAMP} connects the results of this section to the problem of tuning the threshold parameters in AMP. 

\subsection{Optimizing the ideal risk}\label{sec:gdidealrisk}
Let $\tilde{x} \in \mathbb{R}^N$ denote a noisy observation of the vector $x_o$, i.e., $\tilde{x} = x_o+ w$, where $w \sim N(0,  \sigma^2  I)$. Further assume that the noise variance $\sigma^2$ is known. Since $x_o$ is either a sparse or approximately sparse vector, we can employ soft thresholding function to obtain an estimate of $x_o$:
\[
\hat{x}_{\tau} = \eta (\tilde{x} ; \tau).
\]  
This denoising scheme has been proposed in \cite{donoho1995noising}, and its optimality properties have been studied in the minimax framework. As is clear from the above formulation, the quality of this estimate is determined by the parameter $\tau$. Furthermore, the optimal value of $\tau$ depends both on the signal and on the noise level. Suppose that we consider the MSE to measure the goodness of the estimate $\hat{x}_{\tau}$:
\[
r(\tau) \triangleq \frac{1}{N} \mathbb{E} \|\hat{x}_\tau - x_o\|_2^2. 
\]
According to this criterion, the optimal value of $\tau$ is the one that minimizes $r(\tau)$. For the moment assume that $r(\tau)$ is given and forget the fact that $r(\tau)$ is a function of $x_o$ and hence is not known in practice. Can we find the optimal value of $\tau$ defined as
\begin{equation}\label{eq:optparam}
\tau_{opt} = \arg\min_{\tau} r(\tau)
\end{equation}
efficiently? The following lemma simplifies the answer to this question.

\begin{lemma}\label{lem:qcvx} {\rm \cite{mousavi2013asymptotic}}
$r (\tau)$ is a quasi-convex function of $\tau$. Furthermore, the derivative of the function is equal to zero in at most one finite value of $\tau$ and that is $\tau_{\text{opt}}$.  
\end{lemma}

\begin{figure}[t!]
\centering
\includegraphics[width= 11cm]{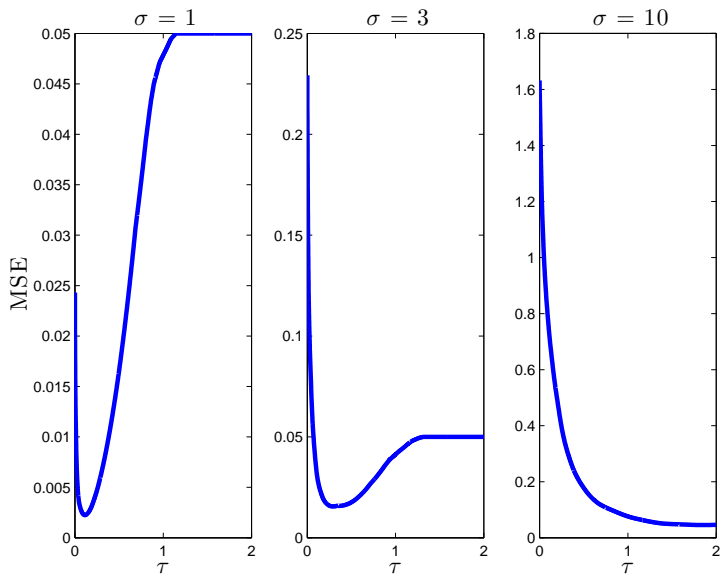}
\caption{Three different forms for MSE vs. $\tau$. Three plots correspond to two different standard deviation of the noise in the observation.}
\label{fig:MSE3}
\end{figure}

In other words, we will in general observe three different forms for $r(\tau)$. These three forms are shown in Figure \ref{fig:MSE3}. Suppose that we aim to obtain $\tau_{\text{opt}}$. Lemma \ref{lem:qcvx} implies that the gradient of $r(\tau)$ at any $\tau$ points toward $\tau_{opt}$. Therefore, we expect the gradient descent algorithm to converge to $\tau_{opt}$. Let $\gamma_t$ denote the estimate of the gradient descent algorithm at iteration $t$. Then, the updates of the algorithm are given by
\begin{align}\label{eq:GDmain}
\gamma_{t+1} = \gamma_{t} - \alpha  \frac{dr(\gamma_t)}{d \tau}, 
\end{align}
where $\alpha$ is the step size parameter. For instance, if $L$ is an upped bound on the second derivative of $r(\tau)$, then we can set $\alpha = 1/L$.\footnote{In practice, we employ back-tracking to set the step-size. } Our first result shows that, even though the function is not convex, the gradient descent algorithm converges to the optimal value of $\tau$. 

\begin{lemma}\label{lem:GD}
Let $\alpha = \frac{1}{L}$ and suppose that the optimizing $\tau$ is finite. Then, $\lim_{t \rightarrow \infty} \frac{ dr (\gamma_t)}{d \tau} = 0$.  
\end{lemma}
See Section \ref{subset:GD} for the proof of this lemma. Note that the properties of the risk function summarized in Lemma \ref{lem:qcvx} enable us to employ standard techniques to prove the convergence of \eqref{eq:GDmain}. 

The discussions above are useful if the risk function and its derivative are given. But these two quantities are usually not known in practice. Hence we need to estimate them. The next section explains how we estimate these two quantities. 

\subsection{Approximate gradient descent algorithm}\label{ssec:empgraddes}

In Section \ref{ssec:tuneint} we described a method to estimate the risk of the soft thresholding function. Here we formally define this empirical unbiased estimate of the risk in the following way:

\begin{definition}\label{def:EmpEst}
The empirical unbiased estimate of the risk is defined as
\begin{equation}
\hat{r}(\tau) \triangleq  \frac{1}{N}\|\eta(\tilde{x};\tau) - \tilde{x} \|_2^2-  \sigma^2 + 2 \sigma^2 (\mathbf{1}^T ( \eta'(\tilde{x}; \tau)-\mathbf{1}))
\end{equation}
\end{definition}
Here, for notational simplicity, we assume that the variance of the noise is given. In Section \ref{sec:simul} we show that estimating $\sigma$ is straightforward for AMP. Instead of estimating the optimal parameter $\tau_{opt}$ through \eqref{eq:optparam}, one may employ the following optimization:
\begin{align}\label{equ:EstOptTau}
\hat{\tau}_{\rm opt} \triangleq \arg \min_{\tau} \hat{r}(\tau).
\end{align}
This approach was proposed by Donoho and Johnstone \cite{donoho1994threshold}, and the properties of this estimator are derived in \cite{donoho1995adapting}. However, \cite{donoho1995adapting} does not provide an algorithm for finding $\hat{\tau}_{opt}$. Exhaustive search approaches are computationally very demanding and hence not very useful for practical purposes.\footnote{Note that $\tau_{opt}$ must be estimated at every iteration of AMP. Hence we seek very efficient algorithms for this purpose. } As discussed in Section \ref{sec:gdidealrisk}, one approach to reduce the computational complexity is to use the gradient descent algorithm. Needless to say that the gradient of $r(\tau)$ is not given, and hence it has to be estimated. One simple idea to estimate the gradient of $r(\tau)$ is the following: Fix $\Delta_N$ and estimate the derivative according to
\begin{align}\label{equ:der}
\frac{d \hat{r}(\tau)}{d \tau} = \frac{\hat{r} (\tau+ \Delta_N) - \hat{r}(\tau)}{\Delta_N}. 
\end{align}
We will prove in Section \ref{ssec:derivaccur} that, if $\Delta_N$ is chosen properly, then as $N \rightarrow \infty$, $\frac{\hat{dr(\tau)}}{d \tau}  \rightarrow \frac{dr(\tau)}{d \tau}$ in probability. Therefore, intuitively speaking, if we plug in the estimate of the gradient in \eqref{eq:GDmain}, the resulting algorithm will perform well for large values of $N$. We will prove in the next section that this intuition is in fact true. Note that since we have introduced $\Delta_N$ in the algorithm, it is not completely free of parameters. However, we will show both theoretically and empirically, the performance of the algorithm is not sensitive to the actual value of $\Delta_N$. Hence, the problem of setting $\Delta_N$ is simple and inspired by our theoretical results we will provide suggestions for the value of this parameter in Section \ref{sec:simul}. 

\begin{figure}[t!]
\centering
\includegraphics[width= 12cm]{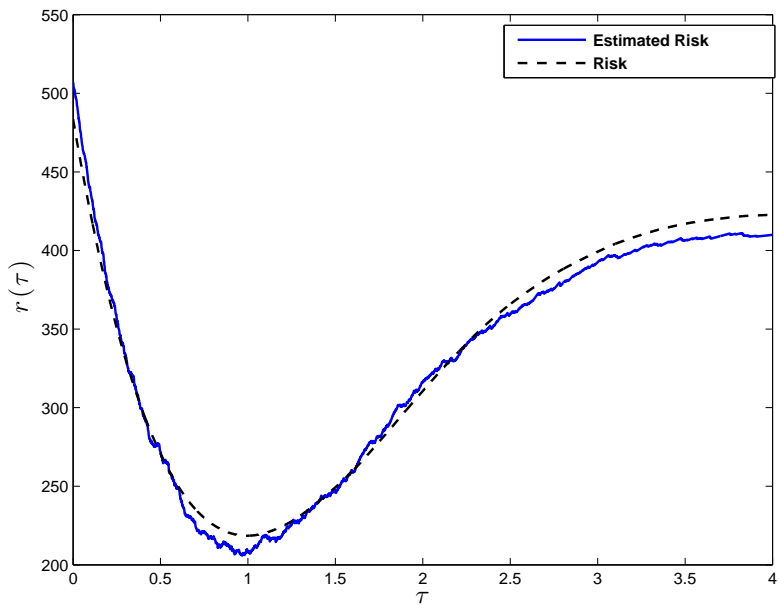}
\caption{The dashed black curve denotes the risk function and the solid blue curve indicates its estimation. For the model we have used in order to produce this plot refer to Section \ref{sec:simul}. Measurements are noiseless.} 
\label{fig:diff_est_real}
\end{figure}

Therefore, our approximate gradient descent algorithm uses the following iteration:
\begin{eqnarray}
\tau^{t+1} = \tau^t- \alpha \frac{d \hat{r}(\tau^t)}{d \tau},
\end{eqnarray}
where as before $\tau^t$ is the estimate of $\tau_{opt}$ at iteration $t$ and $\alpha$ denotes the step size. Before, we proceed to the analysis section, let us clarify some of the issues that may cause problem for our approximate gradient descent algorithm. First note that since $\hat{r}(\tau)$ is an estimate of $r(\tau)$, it is not a quasi-convex any more. Figure \ref{fig:diff_est_real} compares $r(\tau)$ and $\hat{r} (\tau)$.  As is clear from this figure $\hat{r} (\tau)$ may have more than one local minima. One important challenge is to ensure that our algorithm is trapped in a local minima that is ``close to'' the global minima of $r(\tau)$. We will address this issue in Section \ref{sec:graddesth}.

\subsection{Accuracy of the gradient descent algorithm}

\subsubsection{Our approach}
The goal of this section is to provide performance guarantees for the empirical gradient descent algorithm that is described in Section \ref{ssec:empgraddes}. We achieve this goal in three steps: (i) characterizing the accuracy of the empirical unbiased risk estimate $\hat{r}(\tau)$ in Section \ref{sec:empriskth}, (ii) characterizing the accuracy of the empirical estimate of the derivative of the risk $\frac{d \hat{r}}{d \tau}$ in Section \ref{ssec:derivaccur}, and finally (iii) providing a performance guarantee for the approximate gradient descent algorithm in Section \ref{sec:graddesth}.

\subsubsection{Accuracy of empirical risk}\label{sec:empriskth}

Our first result is concerned with the accuracy of the risk estimate $\hat{r}(\tau)$. Consider the following assumption: we know a value $\tau_{max}$, where $\tau_{opt} < \tau_{max}$.\footnote{Note that this is not a major loss of generality, since $\tau_{max}$ can be as large as we require. }

\begin{theorem}\label{thm:bound}
Let $r(\tau)$ be defined according to \eqref{eq:stein} and $\hat{r}(\tau)$  be as defined in Definition \ref{def:EmpEst}. Then,
\begin{align}
&\mathbb{P} \Big(\sup_{ 0< \tau < \tau_{max}} |r(\tau) - \hat{r}(\tau)| \geq (2+ 4 \tau_{\max})N^{-1/2+ \epsilon} \Big) \nonumber \\
 &\leq 2N e^{-2N^{2 \epsilon}}+ 2\tau_{\max}^2N^{\frac{3}{2}-\epsilon} e^{-\frac{2c^2N^{2\epsilon}}{\tau_{\max}^2}} \nonumber,
\end{align}
where $\epsilon < 1/2$ is an arbitrary but fixed number. 
\end{theorem}
See Section \ref{subset:bound} for the proof of Theorem \ref{thm:bound}. First note that the probability on the right hand side goes to zero as $N \rightarrow \infty$. Therefore, we can conclude that according to  Theorem \ref{thm:bound} the difference between $r(\tau)$ and $\hat{r}(\tau)$ is negligible when $N$ is large(with very high probability). Let $\tau_{\text{opt}} = \arg\min_\tau r(\tau)$ and $\hat{\tau}_{\text{opt}} = \arg\min_{\tau} \hat{r}(\tau)$. The following simple corollary of Theorem \eqref{thm:bound} shows that even if we minimize $\hat{r}(\tau)$ instead of $r(\tau)$, still $r(\hat{\tau}_{opt})$ is close to $r(\tau_{opt})$.

\begin{corollary}\label{cor:diff} Let $\tau_{opt}$ and $\hat{\tau}_{opt}$ denote the optimal parameters derived from the actual and empirical risks respectively. Then,
\begin{align}
&\bP\left(|r(\tau_{\text{opt}}) - r(\hat{\tau}_{\text{opt}})| > (4+8 \tau_{\max})N^{-1/2+ \epsilon} \right) \nonumber \\ 
& \leq 2N e^{-2N^{2 \epsilon}}+ 2\tau_{\max}^2N^{\frac{3}{2}-\epsilon} e^{-\frac{2c^2N^{2\epsilon}}{\tau_{\max}^2}}. \nonumber
\end{align}
\end{corollary}
See Section \ref{subsec:diff} for the proof of Corollary \ref{cor:diff}. Corollary \ref{cor:diff} shows that if we could find the global minimizer of the empirical risk, it provides a good estimate for $\tau_{opt}$ for high dimensional problems. The only limitation of this result is that finding the global minimizer of $\hat{r}(\tau)$ is computationally demanding as it requires exhaustive search. Therefore, in the next sections we analyze the fixed points of the approximate gradient descent algorithm.

\subsubsection{Accuracy of the derivative of empirical risk}\label{ssec:derivaccur}
Our next step is to prove that our estimate of the gradient is also accurate when $N$ is large. The estimate of the gradient of $r(\tau)$ is given by 
\begin{align}\label{equ:der1}
\frac{d\hat{r}}{d \tau} = \frac{\hat{r} (\tau+ \Delta_N) - \hat{r}(\tau)}{\Delta_N}. 
\end{align}
The following theorem describes the accuracy of this estimate:

\begin{theorem}\label{thm:der}
Let $\Delta_N = \omega(N^{-1/2+ \epsilon})$ and $\Delta_N = o(1)$ simultaneously. Then, there exists $\tau' \in (\tau, \tau+ \Delta_N)$ such that
\begin{align}
&\mathbb{P} \left(\left| \frac{d\hat{r}}{d \tau} - \frac{d r }{d\tau}(\tau') \right| \geq (8+16 \tau_{\max})N^{-1/2+ \epsilon} \Delta_N^{-1}\right) \nonumber \\
&\leq 2N e^{-2N^{2 \epsilon}}+ 2\tau_{\max}^2N^{\frac{3}{2}-\epsilon} e^{-\frac{2c^2N^{2\epsilon}}{\tau_{\max}^2}}. \nonumber 
\end{align}
In particular, as $N \rightarrow \infty$ $ \frac{d\hat{r}}{d \tau}$ converges to $ \frac{d r }{d\tau}$ in probability. 
\end{theorem}
The proof of Theorem \ref{thm:der} is available in Section \ref{subset:der}. The following remarks highlight some of the main implications of Theorem \ref{thm:der}. \\

\noindent \textbf{Remark:} The difference between the actual derivative of the risk and the estimated one is small for large values of $N$. Therefore, if the actual derivative is positive (and not too small) then the estimated derivative remains positive, and if the actual derivative is negative (and not too small), then the estimated derivative will also be negative. This feature enables the gradient descent with an estimate of the derivative to converge to a point that is close to $\tau_{opt}$. \\

\begin{figure}[t!]
\centering
\includegraphics[width= 9cm]{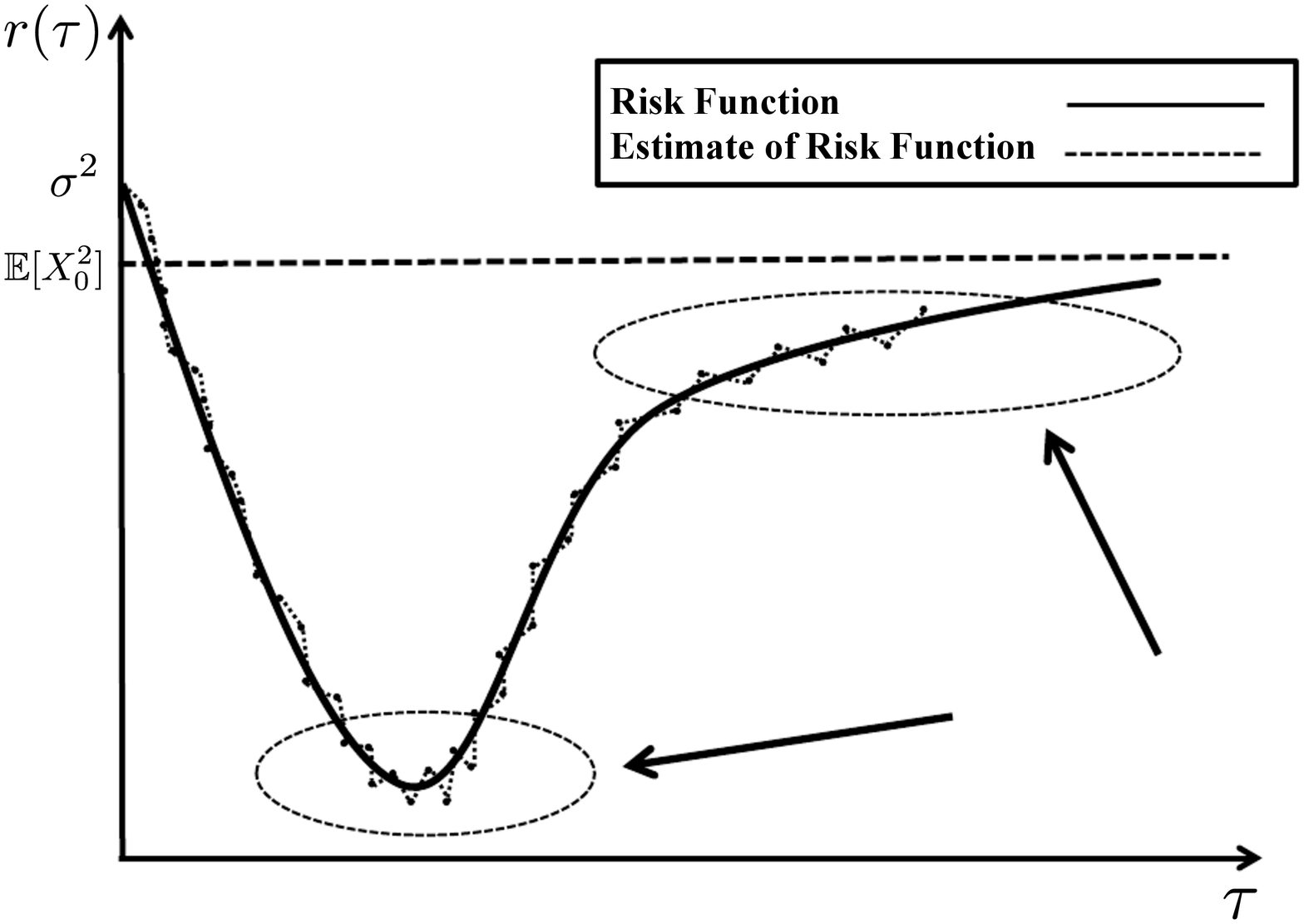}
\caption{Risk function and its estimate. The estimation of the risk function has local minima for the points where $|\frac{d r}{d \tau}(\tau)|=\cO\left(N^{-\frac{1}{2}+\epsilon}\Delta_N^{-1}\right)$. The two regions for which this phenomenon can happen are specified by ellipsoids.} 
\label{fig:riskEst}
\end{figure}

\noindent \textbf{Remark:} Note that the small error that we have in the estimate of the derivative may cause difficulties at the places where the derivative is small. There are two regions for which the derivative is small. As shown in Figure \ref{fig:riskEst}, the first region is around the optimal value of $\tau_{opt}$, and the second region is for very large values of $\tau$. Note that the small error of the estimates  may lead to local minimas in these two regions. We show how the algorithm will avoid the local minimas that occur for large values of $\tau$. Furthermore, we will show that all the local minmas that occur around $\tau_{opt}$ have risk which is close to optimal risk.  \\

\subsubsection{Accuracy of empirical gradient descent}\label{sec:graddesth}

In order to prove the convergence of the gradient descent algorithm we require two assumptions:
\begin{enumerate}
\item[(i)] We know a value $\tau_{max}$, where $\tau_{opt} < \tau_{max}$. 

\item[(ii)] The magnitude of second derivative of $r(\tau)$ is bounded from above by $L$ and $L$ is known.
\end{enumerate}
Before we proceed further let us describe why these two assumptions are required. Note from Figure \ref{fig:riskEst} that for very large values of $\tau$, where the derivative of the ideal risk is close to zero, the empirical risk may have many local minima. Therefore, the gradient descent algorithm is not necessarily successful if it goes to this region. Our first condition is to ensure that we are avoiding this region. So, we modify the gradient descent algorithm in a way that if at a certain iteration it returns $\gamma^t > \tau_{max}$, we realize that this is not a correct estimate. The second condition is used to provide a simple way to set the step size in the gradient descent.  

It is standard in convex optimization literature to avoid the second condition by setting the step-size by using the backtracking method. However, for notational simplicity we avoid back-tracking in our theoretical analysis. However, we will employ it in our final implementation of the algorithm. Similarly, the first constraint can be avoided as well. We will propose an approach in the simulation section to avoid the first condition as well.

Let $\tau^{t}$ denote the estimates of the empirical gradient descent algorithm with step size $\alpha = \frac{1}{L}$. Also, let $\gamma^t$ denote the estimates of the gradient descent on the ideal risk function as introduced in \eqref{eq:GDmain}. We can then prove the following.

\begin{theorem}\label{thm:Conv_GD}
For every iteration $t$ we have,
\[
\lim_{N \rightarrow \infty} |\tau^t - \gamma^t| = 0,
\]
in proability. 
\end{theorem}
\noindent See Section \ref{ssc:Conv_GD} for the proof.

\section{Optimal tuning of AMP}\label{sec:optAMP}

\subsection{Formal statement of AMP features}\label{sec:formalstateAMP}
In Section \ref{ssec:ampintui} we claimed that in asymptotic settings the iterations of AMP can be cast as a sequence of denoising problems. The goal of this section is to formally state this result. Toward this goal we start with the formal definition of the asymptotic settings that is adopted from \cite{DoMaMoNSPT, BaMo10}. Let $n,N \rightarrow \infty$ while $\delta = \frac{n}{N}$ is fixed. We write the vectors and matrices as $x_o(N), A(N), y(N)$, and $w(N)$ to emphasize on the ambient dimension of the problem. Note that the dimensions of $A(N)$, $y(N)$, and $w(N)$ all depend on $\delta$ as well.  Therefore, a more appropriate notation is $A(N, \delta)$, $y(N, \delta)$, and $w(N, \delta)$.  However, since our goal is to fix $\delta$ while we increase $N$ we do not include $\delta$ in our notation. 
 
\begin{definition}\label{def:convseq}
A sequences of instances $\{x_o(N), A(N), w(N)\}$ is called a converging sequence if the following conditions hold:
\begin{itemize}
\item[-] The empirical distribution of $x_o(N) \in \mathbb{R}^N$ converges weakly to a probability measure $p_{X}$ with bounded second moment. Furthermore, $\frac{\|x_o(N)\|_2^2}{N}\rightarrow \mathbb{E}(X^2)$, where $X \sim p_X$. 
\item[-] The empirical distribution of $w(N) \in \mathbb{R}^n$ ($n = \delta N$) converges weakly to a probability measure $p_W$ with bounded second moment. Furthermore, $\frac{\|w(N)\|^2}{n} \rightarrow \mathbb{E}(W^2)= \sigma_w^2$ where $W \sim p_W$. 
\item[-] If $\{e_i\}_{i=1}^N$ denotes the standard basis for $\mathbb{R}^N$, then $\max_i \|A(N) e_i \|_2 \rightarrow 1$ and $\min_i \|A(N) e_i \|_2 \rightarrow 1$ as $N \rightarrow \infty$. 
\end{itemize}
\end{definition}

Note the following appealing features of the above definition:

\begin{enumerate}
\item This definition does not impose any constraint on the limiting distributions $p_X$ or $p_W$.  

 \item The last condition is equivalent to saying that all the columns have asymptotically unit $\ell_2$ norm.  

\end{enumerate}

\begin{definition} \label{def:observables}
Let $\{x_o(N), A(N), w(N)\}$ denote a converging sequences of instances. 
Let $x^{t}(N)$ be a sequence of the estimates of AMP at iteration $t$. Consider a function $\psi: \mathbb{R}^2 \rightarrow \mathbb{R}$. An observable $J_{\psi}$ at time $t$ is defined as 
\[
J_{\psi} \left(x_o, x^{t} \right) = \lim_{N \rightarrow \infty} \frac{1}{N} \sum_{i=1}^N \psi \left(x_{o,i}(N), {x}^{t}_{i}(N) \right).
\]
\end{definition}
A popular choice of the $\psi$ function is $\psi_M(u,v)= (u-v)^2$. For this function the observable has the form:
\[
J_{\psi_M} \left(x_o,{x}^{t}\right) \triangleq \lim_{N \rightarrow \infty} \frac{1}{N} \sum_{i=1}^N  \left( x_{o,i}(N)- {x}^{t}_i(N)\right)^2 =  \lim_{N \rightarrow \infty}  \frac{1}{N} \|x_o - {x}^t \|_2^2,
\]
which is the asymptotic MSE. Another example of $\psi$ function is $\psi_D (u,v) = \mathbb{I} (v \neq 0)$, which leads us to
\begin{equation}\label{eq:detobs}
J_{\psi_D} \left(x_o,{x}^{t}\right) \triangleq \lim_{N \rightarrow \infty} \frac{1}{N} \sum_{i=1}^N \mathbb{I} ({x}^{t}_i \neq 0) = \lim_{N \rightarrow \infty} \frac{\|{x}^{t}\|_0}{N}. 
\end{equation}

 The following result, that was conjectured in \cite{DoMaMo09, DoMaMoNSPT} and was finally proved in \cite{BaMo10}, provides a simple description of the almost sure limits of the observables.

\begin{theorem}\label{thm:ampeqpseudo_lip} Consider the converging sequence $\{x_o(N), A(N), w(N)\}$ and let the elements of $A$ be drawn iid from $N(0,1/n)$. Suppose that ${x}^{t}(N)$ is the estimate of AMP at iteration $t$. 
Then for any pseudo-Lipschitz function $\psi : \mathbb{R}^2 \rightarrow \mathbb{R}$
\[
\lim_{N \rightarrow \infty} \frac{1}{N} \sum_i \psi \left({x}^{t}_{i}(N),{x}_{o,i} \right) = E_{X_o,W} \left[\psi(\eta(X_o+ \sigma^t W; \tau^t), X_o)\right]
\] 
almost surely, where on the right hand side $X_o$ and $W$ are two random variables with distributions $p_X$ and $N(0,1)$, respectively. $\sigma^t$ satisifies
\begin{eqnarray} \label{eq:ampevolution}
(\sigma^{t+1})^2 &=& \sigma_{\omega}^2+\frac{1}{\delta} \mathbb{E}_{X, W} \left[(\eta(X + \sigma^t W; \tau^t ) -X)^2\right], \nonumber \\
\sigma_0^2 &=& \frac{\mathbb{E} \left[X_o^2\right]}{\delta}.  
\end{eqnarray}
\end{theorem}

The last two equations are known as {\em state evolution} for the AMP algorithm. According to this theorem as long as the calculation of the pseudo-Lipschitz observables is concerned, we can assume that estimate of the AMP are modeled as iid elements with each element as $\eta (X_{o} + \sigma^{t}W;\tau^t)$ in law, where $X_{o} \sim p_X$ and $W \sim N(0,1)$.  In many cases, it turns out that even if the pseudo-Lipshcitz condition is not satisfied, the signal plus Gaussian noise model still holds. For more information, see \cite{DoMaMoNSPT}.

\subsection{Tuning procedure of AMP}

Inspired by the formulation of AMP, define the following Bayesian risk function for the soft thresholding algorithm:
\begin{align}\label{equ:BayesRisk}
R_{B} (\sigma, \tau ; p_{X}) = \mathbb{E} (\eta(X_o+ \sigma W ; \tau) - X_o)^2,
\end{align}
where the expected value is with respect to two independent random variables $X_o \sim p_X$ and $W \sim N(0,1)$.  One of the main features of this risk function is the following
 \begin{lemma} \label{lem:monotone}
 $ R_{B} (\sigma, \tau; p_X) $ is an increasing function of $\sigma$. 
 \end{lemma}

While this result is quite intuitive and simple to prove, it has an important implication for the AMP algorithm. Let $\tau^1, \tau^2, \ldots$ denote the thresholds of the AMP algorithm at  iterations $t=1, 2, \ldots $. Clearly, the variance of the noise $\sigma$ at iteration $T$ depends on all the thresholds $\tau^1, \tau^2, \ldots, \tau^{T}$ (See Theorem \ref{thm:ampeqpseudo_lip} for the definition of $\sigma$). Therefore, consider the notation $\sigma^{t} (\tau^1, \tau^2, \ldots, \tau^{t})$ for the value of $\sigma$ at iteration $t$. 

\begin{definition}\label{def:tau}
A sequence of threshold parameters $\tau^{*,1}, \tau^{*,2}, \ldots, \tau^{*,T}$ is called optimal for iteration $T$, if and only if
\[
\sigma^t (\tau^{*,1}, \ldots, \tau^{*,T}) \leq \sigma^{t} (\tau^1, \tau^2, \ldots, \tau^T), \ \ \  \forall \tau^1, \tau^2, \ldots, \tau^T \in [0, \infty)^T.
\] 
\end{definition}
Note that in the above definition we have assumed that the optimal value of $\sigma^t$ is achieved by $(\tau^{*,1}, \ldots, \tau^{*,T})$. This assumption is violated for the case $X_o =0$. While we can generalize the definition to include this case, for notational simplicity we skip this special case. The optimal sequence of thresholds has the following two properties:
\begin{enumerate}
\item It provides the fastest convergence rate for the $T^{\rm th}$ iteration. 
\item If we plan to stop the algorithm after $T$ iterations, then it gives the best achievable MSE.\end{enumerate}
These two claims will be clarified as we proceed. According to Definition  \ref{def:tau}, it seems that, in order to tune AMP optimally, we need to know the number of iterations we plan to run it. However, this is not the case for AMP. In fact, at each step of the AMP, we can optimize the threshold as if we plan to stop the algorithm in the next iteration. The resulting sequence of thresholds will be optimal for any iteration $T$. The following theorem formally states this result. 

\begin{theorem}\label{thm:stepwiseoptimal}
Let $\tau^{*,1}, \tau^{*,2}, \ldots, \tau^{*,T}$ be optimal for iteration $T$. Then, $\tau^{*,1}$, \newline $\tau^{*,2}$, $\ldots, \tau^{*,t}$ is optimal for any iteration $t < T$. 
\end{theorem}
See Section \ref{sec:prooftheoremgreedyoptimal} for the proof of this result.  This theorem, while it is simple to prove, provides a connection between optimizing the parameters of AMP and the optimal parameter tuning we discussed for the soft thresholding function. For instance, a special case of the above theorem implies that $\tau^{*,1}$ must be optimal for the first iteration. Intuitively speaking, the signal plus Gaussian noise model is correct for this iteration. Hence we can apply the approximate gradient descent algorithm to obtain $\tau^{*,1}$. Once $\tau^{*,1}$ is calculated we calculate $\tilde{x}^2$ and again from the above theorem we know that $\tau^{*,2}$ should be optimal for the denoising problem we obtain in this step. Therefore, we apply approximate gradient descent to obtain an estimate of $\tau^{*,2}$. We continue this process until the algorithm converges to the right solution. 
 
If we have access to the risk function, then the above procedure can be applied. At every iteration, we find the optimal parameter with the strategy described in Section \ref{sec:gdidealrisk} and the resulting algorithm is optimal for any iteration $t$. However, as we discussed before the risk function is not available. Hence we have to estimate it. Once we estimate the risk function, we can employ the approximate gradient descent strategy described in Section \ref{sec:gdidealrisk}. Consider the following risk estimate that is inspired by SURE: 

\begin{align}\label{eq:empriskdefamp}
\frac{\hat{r}^t(\tau^t)}{N}&=\frac{1}{N} \|\eta(x^t+A^*z^t;\tau^t)-(x^t+A^*z^t)\|_2^2+\left(\sigma^t\right)^2\nonumber \\
&~~~~+\frac{1}{N}2\left(\sigma^t\right)^2\left[ \mathbf{1}^T(\eta'(x^t+A^*z^t;\tau^t)-\mathbf{1})\right].
\end{align}
As is clear from our discussion about the soft thresholding function in Section \ref{ssec:empgraddes}, we would like to apply the approximate gradient descent algorithm to $\frac{\hat{r}(\tau^t)}{N}$. Nevertheless, the question we have to address is that if it is really going to converge to $R_{B} (\sigma, \tau ; p_{X}) $? The next theorem establishes this result.
\begin{theorem}\label{thm:amptune} Let $\frac{\hat{r}^t(\tau^t)}{N}$ denote the estimate of the risk at iteration $t$ of AMP as defined in \eqref{eq:empriskdefamp}. Then,
\begin{align}\label{eq:MSEAMP}
\lim_{N\rightarrow \infty} \frac{\hat{r}^{t}(\tau^t)}{N} = \bE_{X_o,W}\left[(\eta(X_o+\sigma^tW;\tau^t)-X_o)^2\right],
\end{align}
almost surely, where $X_o$ and $W$ are two random variables with distributions $p_X$ and $N(0,1)$, respectively and $\sigma^t$ satisifies \eqref{eq:ampevolution}.
\end{theorem}
\noindent See Section \ref{subsec:amptune} for the proof of this theorem. This result justifies the application of the approximate gradient descent for the iterations of AMP. However, as we discussed in Section \ref{subsec:tune} a rigorous proof of the accuracy of approximate gradient descent requires a stronger notion of convergence. Hence, the result of Theorem \ref{thm:amptune} is not sufficient. One sufficient condition is stated in the next theorem. Let $\hat{\tau}^{t, s}$ denote the estimate of the $s^{\rm th}$ iteration of approximate gradient descent algorithm at the $t^{\rm th}$ iteration of AMP. In addition, let $\gamma^{t,s}$ denote the estimate of the gradient descent algorithm on the ideal risk at the $t^{th}$ iteration of AMP. 

\begin{theorem}
Suppose that there exists $\epsilon>0$ such that for the $t^{\rm th}$ iteration of AMP we have
\begin{equation}\label{eq:requiredcondition}
\mathbb{P} \left( \sup_{\tau^t} \left|\frac{\hat{r}^t(\tau^t)}{N} - \bE_{X_o,W}\left[(\eta(X_o+\sigma^tW;\tau^t)-X_o)^2\right] \right| > cN^{-\epsilon} \right) \rightarrow 0,
\end{equation}
as $N \rightarrow \infty$. If $\Delta_N = N^{-\epsilon/2}$, then
\[
\lim_{N \rightarrow \infty} |\hat{\tau}^{t, s} - \gamma^{t,s}| =0
\]
in proability.
\end{theorem}
The proof of this result is a combination of the proofs of Theorems \ref{thm:der} and \ref{thm:Conv_GD} and is omitted.  Note that \eqref{eq:requiredcondition} has not been proved for the iterations of AMP and remains an open problem.  

\section{Proofs of the main results}\label{sec:proofs}
This section includes the proofs of the results we have unveiled in Sections \ref{subsec:tune} and \ref{sec:optAMP}. 
\subsection{Proof of Lemma \ref{lem:GD} }\label{subset:GD}
In order to prove this lemma we use the procedure introduced in \cite{nesterov2004introductory}. Let $L$ be a constant such that $\left |\frac{d^2 r(\tau)}{d \tau^2}\right | \leq L$. Therefore, according to Lemma 1.2.3 of \cite{nesterov2004introductory}, we can write
\begin{align}\label{eq:GDin}
r(\gamma_{t+1}) \leq r(\gamma_{t}) + \frac{d r}{d \tau}(\gamma_{t})(\gamma_{t+1}-\gamma_{t})+\frac{L}{2}(\gamma_{t+1}-\gamma_{t})^2.
\end{align}
Applying \eqref{eq:GDmain} in \eqref{eq:GDin} yields
\begin{align}\label{eq:GDin2}
r(\gamma_{t+1}) &\leq r(\gamma_{t})-\alpha\left(\frac{d r}{d \tau}(\gamma_{t})\right)^2+\frac{\alpha^2 L}{2} \left(\frac{d r}{d \tau}(\gamma_{t})\right)^2 \nonumber \\
&=r(\gamma_{t})-\left(\frac{\alpha^2 L}{2}-\alpha\right) \left(\frac{d r}{d \tau}(\gamma_{t})\right)^2.
\end{align}
Minimizing the RHS of \eqref{eq:GDin2} with respect to $\alpha$ gives $\alpha=\frac{1}{L}$, and as a result we can write 
\begin{align}\label{eq:GDin3}
r(\gamma_{t+1}) \leq r(\gamma_{t}) -\frac{1}{2L}\left(\frac{d r}{d \tau}(\gamma_{t})\right)^2.
\end{align}
Equation \eqref{eq:GDin3} shows the amount of the decrease of risk at every iteration.
It is straightforward to conclude from \eqref{eq:GDin2} that
\[
r(\gamma_0)- r(\gamma_t ) \geq  \sum_{i=1}^t \frac{1}{2L}\left(\frac{d r}{d \tau}(\gamma_{i})\right)^2.
\]
Combined with the fact that $r(\gamma_t) > r(\tau^*)$, we obtain
\[
r(\gamma_0)- r(\tau^* ) \geq  \sum_{i=1}^t \frac{1}{2L}\left(\frac{d r}{d \tau}(\gamma_{i})\right)^2.
\]
Therefore, it is clear that as $t \rightarrow \infty$, we have $\frac{d r}{d \tau}(\gamma_{t}) \rightarrow 0$. 
 Since $\frac{d r (\tau)}{d \tau}$ is 0 only at $\tau_{\text{opt}}$, we conclude that  $\lim_{t \rightarrow \infty} \gamma_t = \tau_{\text{opt}}$. 

\subsection{Proof of Theorem \ref{thm:bound}}\label{subset:bound}
According to Definition \ref{def:EmpEst}, 
\begin{align}
\hat{r}(\tau) &=  \frac{1}{N} \|\eta(\tilde{x}; \tau) - \nu\|_2^2-  \sigma^2 +  \frac{2 \sigma^2}{N}  (\mathbf{1}^T (\eta'(\tilde{x}; \tau)-\mathbf{1})) \nonumber \\
&=\underbrace{ \frac{1}{N}\sum_{i=1}^N (\eta(\tilde{x}_i; \tau) - \tilde{x}_i)^2}_{\Delta_1(\tau)}-\sigma^2+\underbrace{\frac{2 \sigma^2}{N} \sum_{i=1}^N (\eta'(\tilde{x}_i; \tau)-1)}_{\Delta_2(\tau)}.
\end{align}
Note that, from Lemma \ref{lem:stein}, we conclude that $\mathbb{E} (\hat{r} (\tau)) = r(\tau)$. Since we assume that $\sigma^2$ is known, we should only prove that $\Delta_1$ and $\Delta_2$ are ``close to'' $\bE[\Delta_1]$ and $\bE[\Delta_2]$, respectively. Intuitively speaking, this seems to be correct, as both $\Delta_1(\tau)$ and $\Delta_2(\tau)$ are the empirical averages of $N$ independent samples. The main challenge is the fact that we are interested in the uniform bounds, i.e., bounds that hold for every value of $\tau$. In particular, we would like to show that 
\begin{align}
&\mathbb{P} \left( \sup_{\tau < \tau_{max}} |\Delta_1(\tau) - \mathbb{E} (\Delta_1(\tau))| > (1+ 4 \tau_{max})N^{-\frac{1}{2}+\epsilon} \right) \nonumber \\
&\leq 2 \tau_{\max}N^{3/2- \epsilon} e^{-\frac{2 N^{2\epsilon}}{\tau_{\max}^2}}, \nonumber \\
&\mathbb{P} \left(\sup_{\tau< \tau_{\max}} |\Delta_2(\tau) - \mathbb{E} (\Delta_2(\tau))| > N^{-\frac{1}{2}+\epsilon}\right) \nonumber \\ 
&\leq 2N e^{-{2N^{2\epsilon}}}. \nonumber
\end{align}

\begin{itemize}
\item{\textbf{Step I:  Discussion of $\Delta_1(\tau)$}

Considering a specific value of $\tau$, and call it $\Ttau\in [0,\tau_{max})$. Note that $|\eta(\tilde{x}_i ; \tilde{\tau}) - \tilde{x}_i| < \tilde{\tau}$. Furthermore, $\tilde{x}_i$s are independent random variables. Hence we can employ Hoeffding inequality \cite{boucheron2013concentration} to conclude that
\begin{align}\label{eq:Hoeff1}
\bP \left(|\Delta_1(\tilde{\tau}) - \bE[\Delta_1(\Ttau)]| \geq \alpha \right) \leq 2e^{-\frac{2N^2 \alpha^2}{N\Ttau^2}}. 
\end{align}
Let $\epsilon$ be a small positive number, and plug $\alpha=N^{-\frac{1}{2}+\epsilon}$ in \eqref{eq:Hoeff1} to obtain
\begin{align}\label{eq:Hoeff2}
\bP \left(|\Delta_1(\Ttau) - \bE[\Delta_1(\Ttau)]| \geq N^{-\frac{1}{2}+\epsilon} \right) \leq 2e^{-\frac{2N^{2\epsilon}}{\Ttau^2}} \leq 2e^{-\frac{2N^{2\epsilon}}{\tau_{max}^2}} .
\end{align}
This equation ensures that for large values of $N$, $\Delta_1(\tilde{\tau})$ is close to its expected value with high probability. Note that so far we have proved that \eqref{eq:Hoeff2} holds for only one specific value of $\tau$. However, since we are interested in the global behavior of $r(\tau)$, we are interested in the event
\[
\mathcal{E} \triangleq \left\{\sup_{0<\tau< \tau_{\max}} |\Delta_1 - \bE[\Delta_1]| \geq N^{-\frac{1}{2}+ \epsilon} \right\},
\]
and its probability $\mathbb{P} (\mathcal{E})$. Toward this goal, we adopt the following two-step strategy:
\begin{enumerate}
\item Partition the interval $[0, \tau_{\max}]$ to intervals of size $\gamma$ and set $\mathcal{A}_{\gamma} \triangleq \{0, \gamma, 2 \gamma, \ldots, \lceil \frac{\tau_{\max}}{\gamma}\rceil \gamma \}$. $\gamma$ is a parameter that we set later in the proof. The set $A_{\gamma}$ is shown in Figure \ref{fig:lambda}. We first provide an upper bound on the probability of the event
\[
\mathbb{P} ( \sup_{\tau \in \mathcal{A}_{\gamma}} |\Delta_1(\tau) - \mathbb{E} (\Delta_1(\tau))| \geq N^{-1/2+ \epsilon}). 
\]
\item The next step is to provide an upper bound for $|\Delta_1(\tau) - \mathbb{E} (\Delta_1(\tau))|$ for $\tau \in (0, \tau_{max})$ assuming that
\[
|\Delta_1(\tau) - \mathbb{E} (\Delta_1(\tau))| <  N^{-1/2},
\]
 for all the values of $\tau$ in $\mathcal{A}_{\gamma}$.  
\end{enumerate}


The first step is straightforward. Applying the union bound over $\mathcal{A}_\tau$ together with \eqref{eq:Hoeff2} would give 
\begin{align}\label{eq:unionDelta1}
 \bP \left( \sup_{\tau \in \mathcal{A}_\gamma}|\Delta_1(\tau) - \bE[\Delta_1(\tau)]|\geq N^{-\frac{1}{2}+\epsilon} \right) \leq \left\lfloor {\frac{\tau_{\max}}{\gamma}} \right\rfloor 2e^{-\frac{2N^{2\epsilon}}{\tau_{\max}^2}}.
\end{align}

\begin{figure}[t!]
\centering
\includegraphics[width= 9cm]{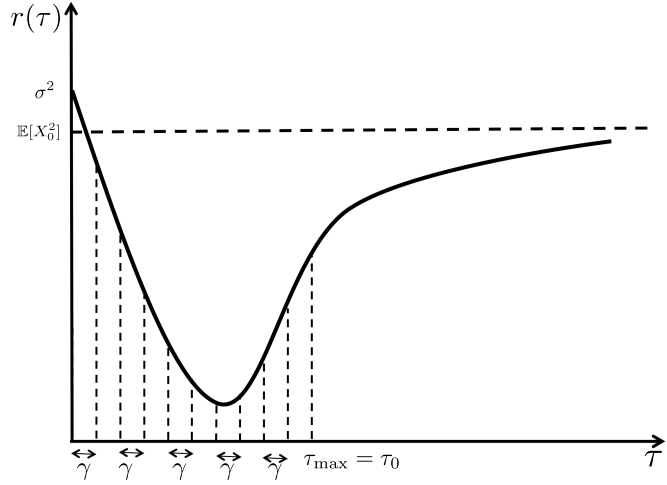}
\caption{Dividing $[0,\tau_{\max}]$ into $\left\lfloor \frac{\tau_{\max}}{\gamma} \right\rfloor$ equally spaced points. We use this procedure to show that $\Delta_1(\tau)$ is ``close to'' $\bE[\Delta_1(\tau)]$ for all $\tau \in [0, \tau_{max}]$. }
\label{fig:lambda}
\end{figure}

As mentioned before the next step of the proof is to provide a bound on the difference $|\Delta_1(\tilde{\tau}) - \mathbb{E} (\Delta_1(\tilde{\tau}))|$ for the values of $\tilde{\tau}$ that are not in $\mathcal{A}_{\gamma}$. Suppose that $\Ttau \in [0,\tau_{\max}]$ and $\Ttau \notin \mathcal{A}_{\gamma}$. Therefore, there exists $\underline{\tau}, \overline{\tau} \in \mathcal{A}_{\gamma}$ such that $\Ttau \in (\underline{\tau},\overline{\tau})$ with $\overline{\tau} - \underline{\tau} < \gamma$. We have
\begin{align}\label{eq:Bounding}
&\Bigg |\sum_{i=1}^N \big(\eta(\tilde{x}_i; \Ttau) - \tilde{x}_i\big)^2- \bE \left[ \sum_{i=1}^N \big(\eta(\tilde{x}_i; \Ttau) - \tilde{x}_i\big)^2 \right] \Bigg |\nonumber \\
& \leq  \Bigg |\sum_{i=1}^N \big( \big| \eta(\tilde{x}_i ; \underline{\tau}) - \tilde{x}_i\big|+\gamma \big)^2 - \bE \left[ \sum_{i=1}^N \big(\big| \eta(\tilde{x}_i; \underline{\tau}) - \tilde{x}_i\big| -\gamma \big)^2 \right] \Bigg | \nonumber \\
& =  \Bigg |\sum_{i=1}^N \big(\eta(\tilde{x}_i ; \underline{\tau}) - \tilde{x}_i\big)^2 - \bE \left[ \sum_{i=1}^N \big(\eta(\tilde{x}_i; \underline{\tau}) - \tilde{x}_i\big)^2 \right] \nonumber \\
&~~~~~~+2\gamma \sum_{i=1}^N \bigg(\big| \eta(\tilde{x}_i ; \underline{\tau}) - \tilde{x}_i\big| + \bE \left[\big| \eta(\tilde{x}_i ; \underline{\tau}) - \tilde{x}_i\big| \right] \bigg) \Bigg | \nonumber \\
&\leq \Bigg |\sum_{i=1}^N \big( \eta(\tilde{x}_i ; \underline{\tau}) - \tilde{x}_i\big)^2 - \bE \left[ \sum_{i=1}^N \big(\eta(\tilde{x}_i ; \underline{\tau}) - \tilde{x}_i\big)^2 \right] \Bigg | +\Bigg | 4\gamma N \tau_{\max} \Bigg |,
\end{align}
where in the last inequality we have applied the triangle inequality along with the fact that $|\eta(\alpha; \tau)-\alpha|\leq \tau \leq \tau_{\max}$.  The last step of bounding $|\Delta_1(\tau)- \mathbb{E} (\Delta_1(\tau))|$ is to set $\gamma$ and combine \eqref{eq:unionDelta1} and \eqref{eq:Bounding}.

If we set $\gamma = \frac{N^{-\frac{3}{2}+\epsilon}}{4 \tau_{\max}}$, then \eqref{eq:unionDelta1} and \eqref{eq:Bounding} prove that
\begin{align}
&\mathbb{P} ( \sup_{\tau < \tau_{max}} |\Delta_1(\tau) - \mathbb{E} (\Delta_1(\tau))| > (1+ 4 \tau_{max})N^{-\frac{1}{2}+\epsilon} ) \nonumber \\ 
&\leq 2 \tau_{\max}N^{3/2- \epsilon} e^{-\frac{2 N^{2\epsilon}}{\tau_{\max}^2}}. \nonumber 
\end{align}
}

\begin{figure}[t!]
\centering
\includegraphics[width= 9cm]{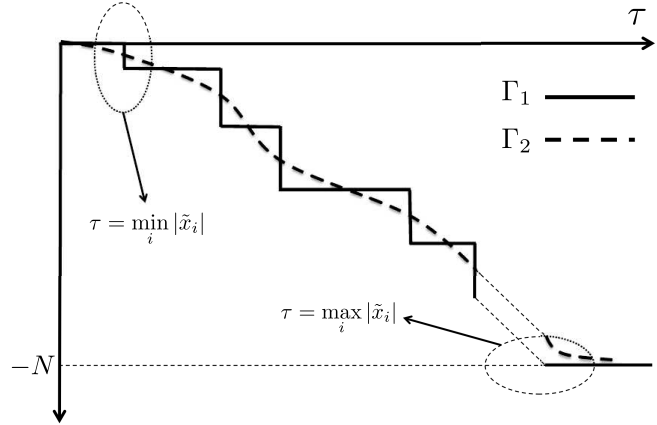}
\caption{Illustration of $\Gamma_1$ and $\Gamma_2$, which are two monotonically decreasing functions with respect to $\tau$. The supremum of the distance between these two functions is achieved at the jump points of $\Gamma_1$, which is peicewise constant.}
\label{fig:diff}
\end{figure}

\item{\textbf{Step II: Discussion of $\Delta_2(\tau)$}

Now we consider $\Delta_2$.  We rewrite $\Delta_2(\tau)-\bE[\Delta_2(\tau)]$ as 
\begin{align}\label{eq:SecDif}
\Delta_2(\tau)-\bE[\Delta_2(\tau)]= \underbrace{ \frac{1}{N}\sum_{i=1}^N (\eta'(\tilde{x}_i ; \tau)-1)}_{\Gamma_1} - \underbrace{ \frac{1}{N} \left(\sum_{i=1}^N \bP(\tilde{x}_i> \tau) -1 \right)}_{\Gamma_2},\end{align}
It is straightforward to show the following two properties for $\Gamma_1$:
\begin{itemize}
\item[(i)] $\Gamma_1$ is a piecewise constant function. The jumps occur at $|\tilde{x}_1|, \ldots, |\tilde{x}_n|$. 

\item[(ii)] $\Gamma_1$ is non-increasing function of $\tau$.
\end{itemize}
This function is exhibited in Figure \ref{fig:diff}. Furthermore, note that $\Gamma_2$ is monotonically decreasing function with respect to $\tau$. Our goal is to bound
\[
\sup_{\tau < \tau_{max}}  |\Gamma_1(\tau)- \Gamma_2(\tau)|. 
\]
However, since $\Gamma_1(\tau)$ is piecewise constant and decreasing this supremum is achieved at one of the jump points $|\tilde{x}_1|, |\tilde{x}_2|, \ldots, |\tilde{x}_n|$, i.e., 
\[
\sup_{\tau < \tau_{max}}  |\Gamma_1(\tau)- \Gamma_2(\tau)| = \sup_{\tau \in \{  |x_{1}| , |x_{2}|, \ldots, |x_{n}|\} }  |\Gamma_1(\tau)- \Gamma_2(\tau)|;
\]
see Figure \ref{fig:diff}. Therefore, by employing the union bound we conclude that
\begin{eqnarray}\label{eq:last1conc}
\lefteqn{\mathbb{P} (\sup_{\tau \in \{  |x_{1}| , |x_{2}|, \ldots, |x_{n}|\} }  |\Gamma_1(\tau)- \Gamma_2(\tau)| > \alpha)} \nonumber \\ 
&\leq& n \mathbb{P} ( |\Gamma_1(\tau)- \Gamma_2(\tau)| > \alpha) \leq 2N {\rm e}^{- {2N \alpha^2}}.
\end{eqnarray}
To obtain the last inequality we have used Hoeffding inequality (Note that $|\eta'(\tilde{x}_i- 1)|<1$). As before we set $\alpha = N^{-1/2+ \epsilon}$ and obtain 
\begin{eqnarray}
\mathbb{P} (\sup_{\tau< \tau_{\max}} |\Delta_2(\tau) - \mathbb{E} (\Delta_2(\tau))| > N^{-\frac{1}{2}+\epsilon}) &\leq& 2N e^{-{2N^{2\epsilon}}}. \nonumber
\end{eqnarray}

}

\end{itemize}

Finally, applying union bound and employing \eqref{eq:unionDelta1} and \eqref{eq:last1conc} completes the proof.

\subsection{Proof of Corollary \ref{cor:diff} }\label{subsec:diff}
Note that, since $\tau_{opt}$ is the minimizer of $r(\tau)$, we have $r(\tau_{opt}) < r (\hat{\tau}_{opt})$. Therefore,
\begin{eqnarray}\label{eq:proofcorempth1}
\lefteqn{r(\hat{\tau}_{opt}) - r(\tau_{opt}) = r(\hat{\tau}_{opt}) - \hat{r} (\hat{\tau}_{opt}) +  \hat{r} (\hat{\tau}_{opt}) -  \hat{r} (\tau_{opt})+  \hat{r} ({\tau}_{opt})  -r(\tau_{opt}) } \nonumber \\
&\leq& r(\hat{\tau}_{opt}) - \hat{r} (\hat{\tau}_{opt}) + \hat{r} ({\tau}_{opt})  -r(\tau_{opt}), \hspace{5cm}
\end{eqnarray}
where the last inequality is due to the fact that $  \hat{r} (\hat{\tau}_{opt}) -  \hat{r} (\tau_{opt}) <0$. Note that according to Theorem \ref{thm:bound} we have
\begin{eqnarray}\label{eq:proofcorempth2}
|r(\hat{\tau}_{opt}) - \hat{r} (\hat{\tau}_{opt})| \leq (2+ 4 \tau_{\max})N^{-1/2+ \epsilon}, \nonumber \\
|\hat{r} ({\tau}_{opt})  -r(\tau_{opt})| \leq (2+ 4 \tau_{\max})N^{-1/2+ \epsilon}.
\end{eqnarray}
with probability at most $2N e^{-2N^{2 \epsilon}}+ 2\tau_{\max}^2N^{\frac{3}{2}-\epsilon} e^{-\frac{2c^2N^{2\epsilon}}{\tau_{\max}^2}}$. Combining \eqref{eq:proofcorempth1} and \eqref{eq:proofcorempth2} completes the proof.

\subsection{Proof of Theorem \ref{thm:der} }\label{subset:der}
First note that
\begin{eqnarray}\label{eq:proofder1}
\lefteqn{\left| \frac{d\hat{r} (\tau)}{d \tau} - \frac{r(\tau+ \Delta_n) - r(\tau)}{\Delta_n} \right| } \nonumber \\
&=&\left| \frac{\hat{r} (\tau+ \Delta_n) - \hat{r}(\tau)}{\Delta_n} -   \frac{r(\tau+ \Delta_n) - r(\tau)}{\Delta_n}   \right| \nonumber\\
& \leq&  \left |\frac{r(\tau+ \Delta_N) - r(\tau+ \Delta_N)}{\Delta_N}\right|+\left |\frac{r(\tau) - r(\tau)}{\Delta_N}\right| \nonumber \\
&\leq& (4+8 \tau_{\max})N^{-1/2+ \epsilon} \Delta_N^{-1},
\end{eqnarray}
where the last inequality is due to Theorem \ref{thm:bound} and holds with probability $1- 2N e^{-2N^{2 \epsilon}}- 2\tau_{\max}^2N^{\frac{3}{2}-\epsilon} e^{-\frac{2c^2N^{2\epsilon}}{\tau_{\max}^2}}$. Furthermore, note that according to the mean value theorem, there exists $\tau' \in [\tau, \tau+\Delta_n]$ for which
\begin{equation}\label{eq:proofder2}
\frac{r(\tau+ \Delta_n) - r(\tau)}{\Delta_n} = \frac{dr(\tau')}{d \tau} \Delta_n. 
\end{equation}
Combining \eqref{eq:proofder1} and \eqref{eq:proofder2} completes the proof.

\subsection{Proof of Theorem \ref{thm:Conv_GD}}\label{ssc:Conv_GD}

The proof is a simple application of what we have proved in the previous sections. We have
\begin{eqnarray}\label{eq:gdproof1}
\tau^{t+1} - \gamma^{t+1} = \tau^t - \gamma^t + \frac{1}{L} \left( \frac{d \hat{r} (\tau^t)}{d \tau} - \frac{d r (\gamma^t) }{d\tau} \right).
\end{eqnarray}
From Theorem \ref{thm:der} we know that, with probability of at least $1-2N e^{-2N^{2 \epsilon}}- 2\tau_{\max}^2N^{\frac{3}{2}-\epsilon} e^{-\frac{2c^2N^{2\epsilon}}{\tau_{\max}^2}}$, there exists $\tau' \in (\tau^t, \tau^t+ \Delta_N)$ such that 
\begin{equation}\label{eq:gdproof2}
\left|\frac{d\hat{r} (\tau^t)}{d \tau} - \frac{d r(\tau')}{d \tau} \right|\leq (8+16 \tau_{\max})N^{-1/2+ \epsilon} \Delta_n^{-1}.  
\end{equation}
Combining \eqref{eq:gdproof1} and \eqref{eq:gdproof2} we obtain
\begin{eqnarray}
\lefteqn{|\tau^{t+1} - \gamma^{t+1}| = \left|\tau^t - \gamma^t + \frac{1}{L} \left( \frac{d \hat{r} (\tau^t)}{d \tau}- \frac{d r(\tau')}{d \tau} + \frac{d r(\tau')}{d \tau}  - \frac{d r (\gamma^t) }{d\tau} \right) \right|} \nonumber \\
&\leq&  |\tau^t - \gamma^t| + \frac{1}{L} \left|\frac{d \hat{r} (\tau^t)}{d \tau}- \frac{d r(\tau')}{d \tau} \right| +\frac{1}{L} \left| \frac{d r(\tau')}{d \tau}  - \frac{d r (\gamma^t) }{d\tau} \right| \nonumber\\
&\leq & |\tau^t - \gamma^t| + N^{-1/2+ \epsilon} \Delta_n^{-1}+ |\tau^t- \gamma^t| = 2 |\tau^t - \gamma^t| +N^{-1/2+ \epsilon} \Delta_n^{-1}.
\end{eqnarray}
The last inequality holds with probability $1-2N e^{-2N^{2 \epsilon}}- 2\tau_{\max}^2N^{\frac{3}{2}-\epsilon} e^{-\frac{2c^2N^{2\epsilon}}{\tau_{\max}^2}}$.
Since we assume that both algorithms start from $0$, it is therefore straightforward to conclude that
\[
|\tau^t - \gamma^t|  \leq 2cN^{-1/2+ \epsilon} \Delta_n^{-1}(1+ 2+ \ldots +2^t) \leq 2^{t+2} cN^{-1/2+ \epsilon}\Delta_n^{-1}, 
\]
with probability greater than $\left( 1-t \left(2N+8\tau_{\max}^2N^{\frac{3}{2}-\epsilon}\right)e^{-\frac{2N^{2\epsilon}\Delta_N^{-2}}{\tau_{\max}^2}}
 \right)$. Letting $N \rightarrow \infty$ completes the proof.

\subsection{Proof of Theorem \ref{thm:amptune} }\label{subsec:amptune}

By applying Stein's lemma to the the RHS of \eqref{eq:MSEAMP}, we can rewrite it as
\begin{align}
&\bE_{X_o,W}\left[(\eta(X_o+\sigma^tW;\tau^t)-X_o)^2\right] \nonumber \\
&~~~=\bE_{X_o,W}\left[(\eta(X_o+\sigma^tW;\tau^t)-(X_o+\sigma^tW))^2\right]+\left(\sigma^t\right)^2\nonumber \\
&~~~~+2\left(\sigma^t\right)^2\bE_{X_o,W}\left[(\eta'(X_o+\sigma^tW;\tau^t)-1)\right].
\end{align}
Similarly, we can decompose the LHS of \eqref{eq:MSEAMP} as 
\begin{align}
\frac{\hat{r}(\tau^t)}{N}&=\frac{1}{N} \|\eta(x^t+A^*z^t;\tau^t)-(x^t+A^*z^t)\|_2^2+\left(\sigma^t\right)^2\nonumber \\
&~~~~+\frac{1}{N}2\left(\sigma^t\right)^2\left[ \mathbf{1}^T(\eta'(x^t+A^*z^t;\tau^t)-\mathbf{1})\right].
\end{align}
Let $A_{:, i}$ denote the $i^{\rm th}$ column of the matrix $A$. Consider the following function
\begin{align}\label{eq:psi1}
\psi_1(x_i^{t+1},x_{o,i})&=\psi_1(\eta(x_i^{t}+A_{(:,i)}^*z^{t},\tau^{t}),x_{o,i}) \nonumber \\
&=\left(\eta(x_i^{t}+A_{(:,i)}^*z^{t},\tau^{t})-(x_i^{t}+A_{(:,i)}^*z^{t})\right)^2.
\end{align}
The RHS of the second equality in \eqref{eq:psi1} consists of the combination of two pseudo-Lipschitz function, namely soft-thresholding and quadratic function, and hence $\psi_1$ is pseudo-Lipschitz itself. Therefore, according to Theorem \ref{thm:ampeqpseudo_lip}, we can write
\begin{align}\label{eq:psi1Asymp}
\lim_{N\rightarrow \infty}\frac{1}{N}\sum_{i=1}^N \psi_1(x_i^{t+1},x_{i,o})&=\bE\left[\psi_1(X_o+\sigma^tW;\tau^t),X_o)\right] \nonumber \\
&=\bE_{X_o,W}\left[(\eta(X_o+\sigma^tW;\tau^t)-(X_o+\sigma^tW))^2\right].
\end{align}
Furthermore, according to Theorem \ref{thm:ampeqpseudo_lip} we have $\lim_{N \rightarrow \infty} \frac{\| {x}^{t+1}(N)\|_0 }{N} = \mathbb{P} (|X_o+ \sigma^t W| \geq \tau^t )$. Consequently, we have
\begin{align}\label{eq:psi2Asymp}
\lim_{N \rightarrow \infty} \frac{\mathbf{1}^T(\eta'(x^t+A^*z^t;\tau^t)-\mathbf{1}) }{N} =\lim_{N \rightarrow \infty} \frac{\| {x}^{t+1}(N)\|_0 }{N} = \mathbb{P} (|X_o+ \sigma^t W| \geq \tau^t ).
\end{align}
Combining \eqref{eq:psi1Asymp} and \eqref{eq:psi2Asymp} establishes
\begin{align}
\lim_{N \rightarrow \infty}\frac{\hat{r}(\tau^t)}{N} = \bE_{X_o,W}\left[(\eta(X_o+\sigma^tW;\tau^t)-X_o)^2\right]. 
\end{align}

\subsection{Proof of Theorem \ref{thm:stepwiseoptimal} } \label{sec:prooftheoremgreedyoptimal}
The proof is by contradiction. 
Suppose that $\tau^{*,1}, \tau^{*, 2}, \ldots, \tau^{*, t}$ are not optimal for iteration $t$. Then there exists $\tau^{1}, \tau^2, \ldots, \tau^{t}$ such that 
\[
\sigma^t (\tau^1, \ldots, \tau^t) < \sigma^t(\tau^{*,1}, \tau^{*, 2}, \ldots, \tau^{*, t}). 
\]
We define the following thresholding policy: $(\tau^1, \ldots, \tau^t,  \tau^{*, t+1}, \ldots, \tau^{*, T})$. We can now prove that
\[
\sigma^{T}(\tau^1, \ldots, \tau^t,  \tau^{*, t+1}, \ldots, \tau^{*, T}) < \sigma^T (\tau^{*,1}, \tau^{*,2}, \ldots, \tau^{*,T}).
\]
The proof is a simple induction. From Theorem \ref{thm:ampeqpseudo_lip} we have
\begin{equation}\label{eq:optimalstepwise}
(\sigma^{t+1})^2 = \sigma_{\omega}^2+\frac{1}{\delta} \mathbb{E}_{X, W} \left[(\eta(X + \sigma^t W; \tau^t ) -X)^2\right].
\end{equation}
Since, $\sigma^t (\tau^1, \ldots, \tau^t) < \sigma^t(\tau^{*,1}, \tau^{*, 2}, \ldots, \tau^{*, t})$, Lemma \ref{lem:monotone} combined with  \eqref{eq:optimalstepwise} prove that 
\[
\sigma^{t+1}(\tau^1, \ldots, \tau^t,  \tau^{*, t+1}) <  \sigma^{t+1} (\tau^{*,1}, \tau^{*,2}, \ldots, \tau^{*,t+1}).
\]
It is clear that by induction we can obtain
\[
\sigma^{T}(\tau^1, \ldots, \tau^t,  \tau^{*, t+1}, \ldots, \tau^{*, T}) < \sigma^T (\tau^{*,1}, \tau^{*,2}, \ldots, \tau^{*,T}),
\]
which contradicts the optimality of $\tau^{*,1}, \tau^{*,2}, \ldots, \tau^{*,T}$.

\section{Simulation results}\label{sec:simul}

The main objective of this section is to evaluate the performance of automatically tuned AMP proposed in Algorithm \ref{alg:GD} via simulations. We specifically discuss the effect of the measurement noise, the choice of parameter $\Delta_N$ used in \eqref{equ:der}, and the impact of the sample size $N$ on the performance of our method.

\subsection{Practical approximate gradient descent algorithm}

The approximate gradient descent algorithm is presented in Algorithm \ref{alg:GD}. We review the main pieces below. 

\subsubsection{Estimating the noise variance}\label{ssc:var}
One of the main assumptions we have made so far is that the variance of the noise is given at every iteration. Needless to say that this is not the case in practical applications and hence we have to estimate the variance of the noise. This problem has been studied elsewhere \cite{MalekiThesis, Mo12Gr}. Here we only mention one of the best approaches that exists for estimating $\sigma^t$. The following estimate 
\[
\left( \hat{\sigma}^t \right)^2 = \frac{1}{n} \sum_{i=1}^n (z^t_i)^2
\]
provides an accurate and unbiased estimate of $\sigma^t$. Here, $z^t_i$ denies the $i^{\rm th}$ coordinate of $z^t$. It can be proved that $\lim_{N \rightarrow \infty} \hat{\sigma}^t = \sigma^2$ \cite{Mo12Gr}.

\subsubsection{Setting the step size}\label{ssc:StepSize}
In Section \ref{sec:graddesth} we discussed the performance of the gradient descent algorithm when the algorithm is employing fixed step size. In practical situations, in most cases one can employ one of the methods that have been developed in the optimization literature for this purpose. Here we employ the simplest of these approaches, i.e., the back-tracking. This scheme is described in Algorithm \ref{alg:GD} along with the other parts of our algorithm. For more information on this algorithm please see \cite{BoydVanderberghe}.

\subsubsection{Avoiding local minima}
As discussed in Section \ref{sec:graddesth}, the approximate gradient descent algorithm may be trapped in two sets of local minima: (i) local minimas that occur at large values of $\tau$, and (ii) local minimas that occur around $\tau_{opt}$. See Figure \ref{fig:riskEst} for more information.  Therefore, it is important to ensure that the algorithm avoids the local minimas that occur for large values of $\tau$. This is straightforward for large values of $N$. Note that for large values of $\tau$, the risk function is equal to $\frac{1}{N} \sum_{i=1}^N |x_{o,i}|^2$, since our estimate is essentially equal to zero. On the other hand, we have an access to the vector $\tilde{x}^t$. If we calculate $\frac{1}{N}\sum_{i=1}^N |x_{o,i} + v_i|^2$, it is straightforward to confirm that as $N \rightarrow \infty$ it converges to $\frac{1}{N} \sum_{i=1}^N |x_{o,i}|^2+ \sigma^2$. Therefore, once we have an estimate of $\sigma^2$, we can easily obtain an estimate for $\mu \triangleq \frac{1}{N} \sum_{i=1}^N |x_{o,i}|^2$:
\[
\hat{\mu} \triangleq \frac{1}{N}\sum_{i=1}^N |\tilde{x}^t_{i}|^2 - \sigma^2.  
\]
If the approximate gradient descent algorithm converges to a value of $\tau$ whose estimated risk is close to $\hat{\mu}$, this indicates that the estimated $\tau$ is not close to $\tau_{opt}$. In this case, we re-run the gradient descent algorithm. To improve the performance of the gradient descent we initialize the algorithm at zero, and reduce the step size. 
It is not difficult to prove that if the step size is small enough and the algorithm is initialized at zero then the algorithm will converge to the correct answer. 

\begin{algorithm}[!t]                 
\caption{Finding the $\arg \min_{\tau} \hat{r}(\tau)$ with the approximate gradient descent algorithm  }         
\label{alg:GD}                     
\begin{algorithmic}                    
\REQUIRE $\hat{r}(\tau),\Delta_N,\kappa,\alpha,\beta,\hat{\mu},m, l_0=20,\text{Flag}=1$
\ENSURE $\arg \min_{\tau} \hat{r}(\tau)$ 
\WHILE{$\text{Flag}=1$}
\STATE $\text{Flag}=0, \tau_{\text{new}}=0$
\FOR {$i=1:m$}
\STATE $\tau_{\text{old}}=\tau_{\text{new}}$
\STATE $\frac{d\hat{r}(\tau_{\text{old}})}{d\tau}=\frac{\hat{r}(\tau_{\text{old}}+\Delta_N)-\hat{r}(\tau_{\text{old}})}{\Delta_N}$
\STATE $\Delta_{\tau}=-\frac{d\hat{r}(\tau_{\text{old}})}{d\tau}$
\STATE $l=l_0$
\WHILE{$\hat{r}(\tau_{\text{old}}+l\Delta_{\tau})>\hat{r}(\tau_{\text{old}})+\alpha l  \Delta_{\tau}\frac{d\hat{r}(\tau_{\text{old}})}{d\tau}$}
\STATE $l=\beta l$
\ENDWHILE
\STATE $\tau_{\text{new}}=\tau_{\text{old}}+l \Delta_{\tau}$
\IF{$\frac{\left| \frac{1}{N}\hat{r}(\tau_{\text{new}})-\hat{\mu}\right|}{\hat{\mu}}<\kappa$ or $\tau_{\text{new}}<0$}
\STATE $l_0 = \frac{l_0}{2}, \text{Flag}=1 $
\STATE break
\ENDIF
\ENDFOR
\ENDWHILE
\end{algorithmic}
\end{algorithm}


\subsection{Accuracy of the approximate gradient descent}

In this section we evaluate the quality of the estimates obtained from approximate gradient descent algorithm at different iterations of AMP. Here the description of our experiment:

\begin{itemize}
\item We set $N=2000$ in all experiments unless we mention otherwise. Number of measurements $n$ and the level of sparsity $k$ are obtained according to $n=\lfloor \delta N \rfloor$ and $k=\lfloor \rho n \rfloor$. In these experiments we set $\delta =0.85$ and $\rho=0.25$. $A$ is a measurement matrix having iid entries drawn from Gaussian distribution $N\left(0,\frac{1}{n}\right)$. The signal to be reconstructed $x_o\in \mathbb{R}^N$ has only $k$ non-zero values. We have tested the performance on several distributions for the non-zero entries of $x_o$. However, in the following experiments we consider a unit point mass at 1 as a distribution for non-zero entries of $x_o$.
\item The maximum number of iterations of AMP is equal to 200. Furthermore,  we let gradient descent to iterate 30 times to find $\hat{\tau}_{opt}$. In most cases, the algorithm converges in less than $10$ iterations. One can study other stopping rules to improve the efficiency of the algorithm. As mentioned on the first page the measurements are given by $y = Ax_o + w$, where $w\sim N(0, \sigma_w^2 I)$. We consider three different cases for the noise: $\sigma = 0, 0.2, 0.4$. 

\item The other parameters of the algorithm are set to $\alpha=0.1$, $\beta=0.3$, and $\kappa=0.05$. Furthermore, $\Delta_N$ (the parameter that is used for estimating the derivative of the risk) is set to $0.05$. It is well-known that the performance of the algorithm is robust to the choice of the parameters $\alpha$ and $\beta$. We will later show that the performance of the algorithm is robust to the choice of $\Delta_N$ as well.
\end{itemize}

\begin{figure}[t!]
\centering
\includegraphics[width= 9.9cm]{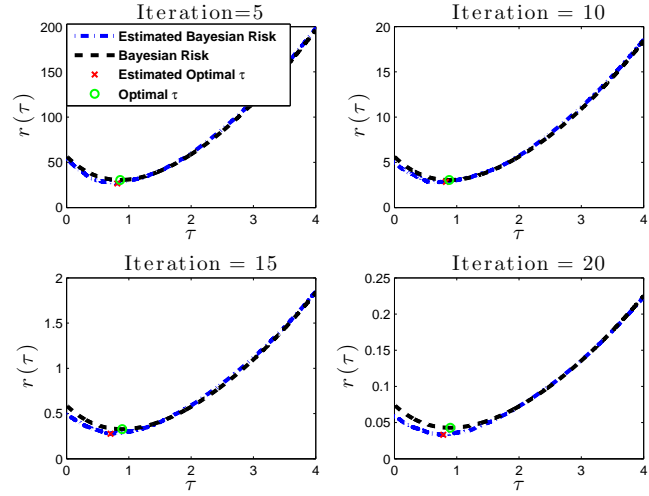}
\caption{Performance of Algorithm \ref{alg:GD} in estimating $\hat{\tau}_{opt}$ in different iterations of AMP. In this experiment $N=2000, \delta=0.85, \rho=0.25$, and we consider noiseless measurements ($\sigma=0$).}
\label{fig:noise_free}
\end{figure}

  Figure \ref{fig:noise_free} shows the performance of Algorithm \ref{alg:GD} for a set of noise-free measurements. It contains 4 plots where they correspond to the different iterations of the AMP. As we mentioned in Section \ref{ssec:tuneint}, at each iteration, AMP calculates $\tilde{x}^t=x_o+v^t$. Each plot in Figure \ref{fig:noise_free} contains the Bayesian risk (defined in \eqref{equ:BayesRisk}), estimate of the Bayesian risk, $\hat{\tau}_{opt}$ and $\tau_{opt}$ defined in \eqref{equ:EstOptTau} and \eqref{eq:optparam}, respectively. 

\begin{figure}[t!]
\centering
\includegraphics[width= 9.9cm]{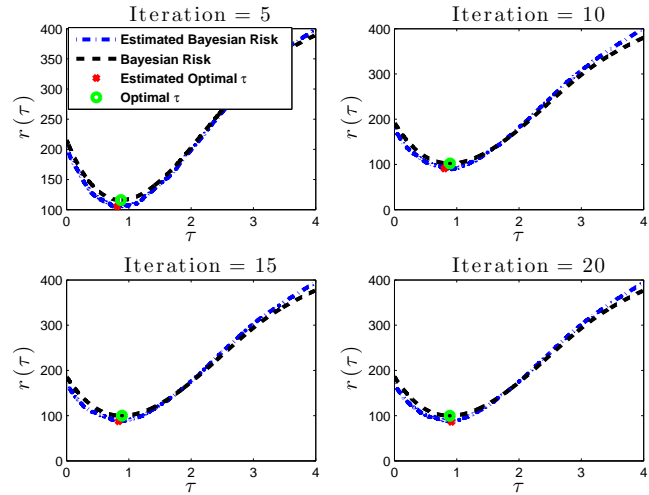}
\caption{Performance of Algorithm \ref{alg:GD} in estimating $\hat{\tau}_{opt}$ in different iterations of AMP. In this experiment $N=2000, \delta=0.85, \rho=0.25$, and the standard deviation of the noise of the measurements $\sigma=0.2$.}
\label{fig:Est_Effect_Iteration_Sig2}
\end{figure}

\begin{figure}[t!]
\centering
\includegraphics[width= 9.9cm]{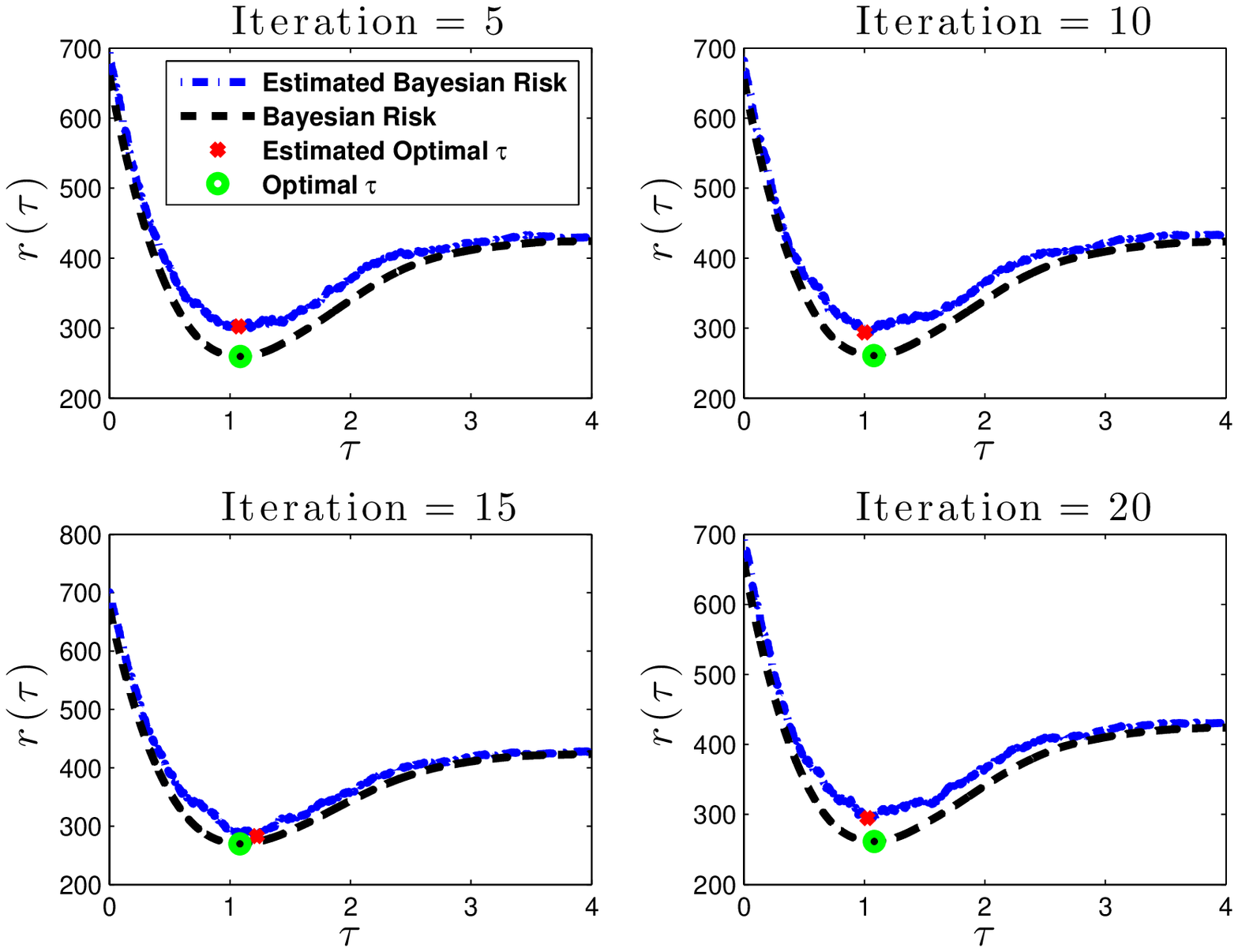}
\caption{Performance of Algorithm \ref{alg:GD} in estimating $\hat{\tau}_{opt}$ in different iterations of AMP. In this experiment $N=2000, \delta=0.85, \rho=0.25$, and the standard deviation of the noise of the measurements $\sigma=0.4$.}
\label{fig:Est_Effect_Iteration_Sig4}
\end{figure}

Figures \ref{fig:Est_Effect_Iteration_Sig2} and \ref{fig:Est_Effect_Iteration_Sig4} are similar to Figure \ref{fig:noise_free} except for the fact that they correspond to the set of noisy measurements with standard deviation 0.2 and 0.4 respectively. As we will show in the next section, the discrepancy between the actual risk and its empirical estimate vanishes as $N$ grows. 



\subsection{Impact of sample size $N$}

\begin{figure}[t!]
\centering
\includegraphics[width= 9.9cm]{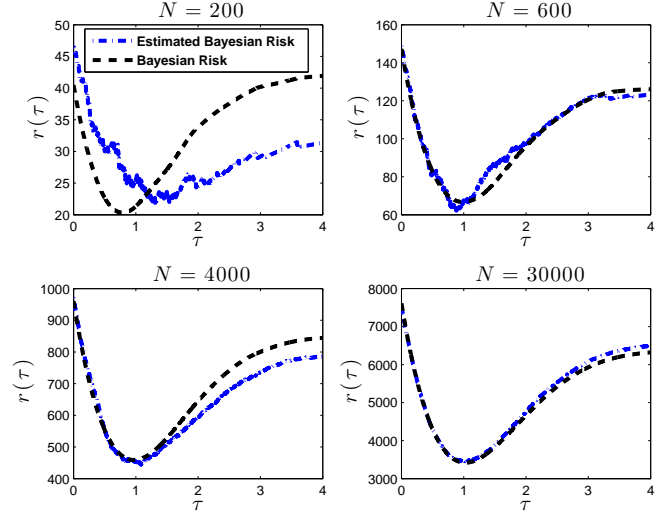}
\caption{Performance of Algorithm \ref{alg:GD} in estimating $\hat{\tau}_{opt}$ for different values of $N$. In this experiment $\delta=0.85, \rho=0.25$, and we consider noiseless measurements ($\sigma=0$).}
\label{fig:Est_Effect_N}
\end{figure}

Next, we test the effect of the sample size $N$ on the accuracy and the performance of the risk estimator. For this experiment, we use noiseless measurements from the same model described at the beginning of this section except for the value of $N$ which is chosen from the set $\{200,600,4000,30000\}$. As we can see in the Figure \ref{fig:Est_Effect_N}, the larger the $N$ the better the performance of the estimator would be. This is exactly what we expect from Throrem \ref{thm:bound}. A rule of thumb that we can give based on many experiments we have done is that, if $N>1000$ and $n > 400$, then the risk estimate is accurate enough and the algorithm will work well. 

\subsection{Setting $\Delta_N$}

\begin{figure}[t!]
\centering
\includegraphics[width= 9.9cm]{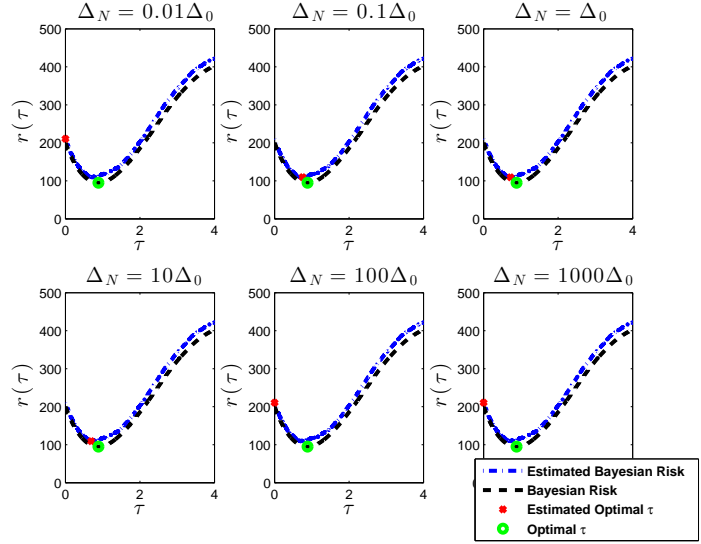}
\caption{Performance of Algorithm \ref{alg:GD} in estimating $\hat{\tau}_{opt}$ for different values of $\Delta_N$. In this experiment $N=2000, \delta=0.85, \rho=0.25$, and the standard deviation of the noise of the measurements $\sigma$ is 0.2. }
\label{fig:EST_Effect_Delta}
\end{figure}

\begin{figure}[t!]
\centering
\includegraphics[width= 9.9cm]{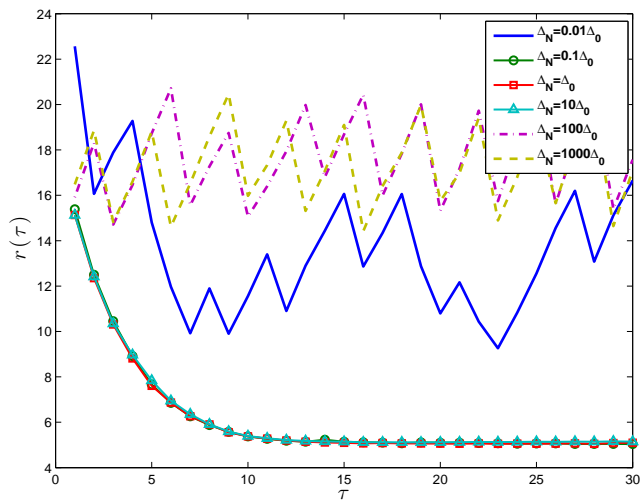}
\caption{Convergence rate and the final MSE for different values of $\Delta_N$. In this experiment $N=2000, \delta=0.85, \rho=0.25$, and the standard deviation of the noise of the measurements $\sigma$ is 0.1.}
\label{fig:MSE_Effect_Delta}
\end{figure}

Figure \ref{fig:EST_Effect_Delta} and Figure \ref{fig:MSE_Effect_Delta} are related to experiment of testing the impact of $\Delta_N$ on our algorithm. We mentioned earlier that $\Delta_N$ is a free parameter; meaning that it could be chosen by the user. Simulation results show that the algorithm is robust to the changes of $\Delta_N$ in a wide range of values. In Algorithm \ref{alg:GD} we fix the $\Delta_N$ as $\Delta_0$ where $\Delta_0=0.05$. Figure \ref{fig:EST_Effect_Delta} shows the performance of Algorithm \ref{alg:GD} in finding the $\hat{\tau}_{opt}$ for different values of $\Delta_N$. In addition, Figure \ref{fig:MSE_Effect_Delta} shows the convergence rate and the final MSE of AMP using different values of $\Delta_N$ in Algorithm \ref{alg:GD}. As we can see from both figures, $\Delta_N$ could be chosen from a wide range of values, from $0.1\Delta_0$ to $10\Delta_0$. As the dimension grows, the range of values for which the algorithm works well expands as well.  In all our experiments $\Delta_N = 0.05$ provides a good value. As $N$ increases, one may choose smaller values of $\Delta$ to have a more accurate the derivative at every iteration. However, this does not provide much improvement in the overall performance of AMP.

\subsection{Comparison with optimal AMP}

\begin{figure}[t!]
\centering
\includegraphics[width= 10cm]{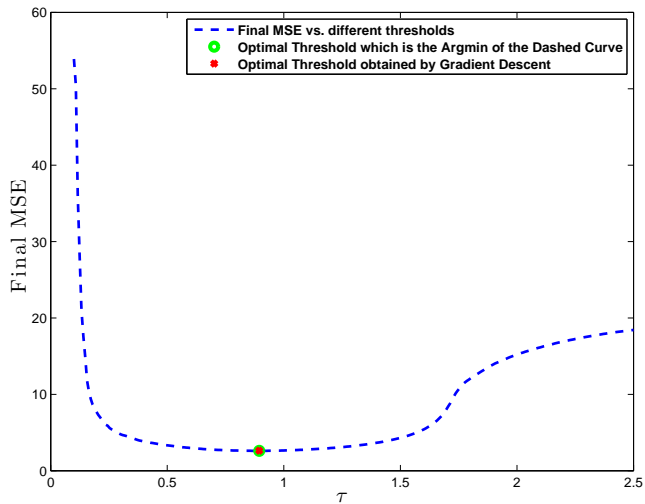}
\caption{Plot of final MSE versus different $\tau \in \mathcal{T}$. We see that the $\tau$ which has the minimum final MSE is very close to the one obtained from tuning the AMP's threshold using the gradient descent method. }
\label{fig:taus}
\end{figure}

Now we compare the performance of our algorithm (AMP with approximate gradient descent (Algorithm \ref{alg:GD})) and optimal AMP derived in \cite{DoMaMoNSPT}. Note that the optimal AMP in \cite{DoMaMoNSPT} requires an oracle information on the sparsity level of the signal to tune the algorithm, while our algorithm is tuned automatically without any information from the user. In this experiment, we consider 300 equally spaced $\tau$ from $[0.1,2.5]$ and call this set $\mathcal{T}$. Each time we fix the $\tau$ (picked up from $\mathcal{T}$) and run the AMP. We name the $\tau$ which gives the minimum final MSE as $\tau_{opt}$. This is the parameter that is also given by the tuning approach in \cite{DoMaMoNSPT}. We also run the AMP using the gradient descent method in Algorithm \ref{alg:GD}. We name the $\tau$ obtained after converging to the final MSE as $\hat{\tau}_{opt}$. Figure \ref{fig:taus} is a comparison of $\tau_{opt}$ and $\hat{\tau}_{opt}$. As we can see, they are very close to each other.

\begin{figure}[t!]
\centering
\includegraphics[width= 10cm]{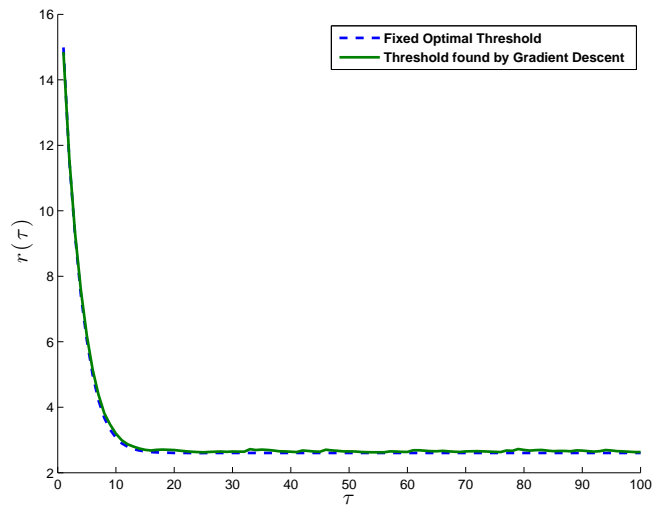}
\caption{Risk of AMP at each iteration for two different approaches. Blue dashed curve shows the MSE for the method in which the threshold is set to the optimal value found from the previous experiment (Figure \ref{fig:taus}). The solid green curve shows the MSE in each iteration when the threshold is set using gradient descent.}
\label{fig:MSE}
\end{figure}

Finally, we compare the convergence of the AMP using the tuning scheme of Algorithm \ref{alg:GD} versus tuning using fixed thresholding policy with $\tau_{opt}$. Figure \ref{fig:MSE} shows the MSE at each iteration. As we can see from the figure, when AMP has converged to the final MSE, performance of the version using fix threshold $\tau_{opt}$ is better than the one which uses gradient descent in each iteration to find the optimal threshold.  

\section{Conclusions}\label{sec:conclusion}
In this paper, we have proposed an automatic approach for tuning the threshold parameters in the AMP algorithm. We proved that (i) this tuning ensures the fastest convergence rate and (ii) The final solution of AMP achieves the minimum mean square reconstruction error that is achievable for AMP. This resolves the problem of tuning the parameters of AMP optimally. 

There are several theoretical and practical problems that are left open for future research. For instance, employing better techniques for estimating the derivative of the risk, and employing better algorithms to employ the approximate derivative to obtain the minimum of the risk function can potentially benefit the performance of the algorithm, Also, it seems that these ideas can be extended to the AMP algorithm for other signal structures.

\bibliographystyle{unsrt}
\bibliography{v7}

\end{document}